\documentclass[revtex4]{emulateapj}
\pdfoutput=1
\usepackage{graphicx} 
\usepackage{subfigure} 
\usepackage{mathtools} 
\usepackage[toc,page]{appendix}
\usepackage{multirow} 
\begin{abstract} 
We investigate the star formation history (SFH) as a function of radius in M101 using archival \textit{HST} ACS photometry.  We derive the SFH from the resolved stellar populations in five $2\arcmin$ wide annuli.  Binning the SFH into time frames corresponding to stellar populations traced by H$\alpha$, far ultraviolet (FUV), and near ultraviolet (NUV) emission, we find that the fraction of stellar populations young enough to contribute in H$\alpha$ is $15\%-35\%$ in the inner regions, compared to less than $5\%$ in the outer regions.  This provides a sufficient explanation for the lack of H$\alpha$ emission at large radii.  We also model the blue to red supergiant ratio in our five annuli, examine the effects that a metallicity gradient and variable SFH have on the predicted ratios, and compare to the observed values.  We find that the radial behavior of our modeled blue to red supergiant ratios is highly sensitive to both spatial variations in the SFH and metallicity.  Incorporating the derived SFH into modeled ratios, we find that we are able to reproduce the observed values at large radii (low metallicity), but at small radii (high metallicity) the modeled and observed ratios are discrepant.

\end{abstract}


\begin{document} \title{The Massive Star Population in M101. II.  Spatial Variations in the Recent Star Formation History} \author{Skyler Grammer and Roberta M. Humphreys} \affil{Minnesota Institute for Astrophysics, 116 Church St SE, University of Minnesota, Minneapolis, MN 55455, USA; grammer@astro.umn.edu, roberta@umn.edu} 

\maketitle

\section{Introduction} 
Observations of spiral galaxies indicate that the stellar populations and emission properties are strongly dependent on radius.  Early observations of the massive star content in M33 demonstrated that the ratio of blue to red supergiants (B/R ratio) decreases with distance from the center \citep{Walker:1964}. Subsequently, many other authors have observed similar phenomena, in a variety of galaxies, which have been attributed to an abundance gradient \citep{Hartwick:1970, Humphreys:1979, Humphreys:1979a, Humphreys:1980, Humphreys:1984, Eggenberger:2002, Grammer:2013a}.  As discussed in \cite{Langer:1995}, and more recently \cite{Meynet:2011}, stellar evolution models are able to reproduce the observed B/R ratios at either high or low metallicity.  However, there is not yet a set of models that reproduce the B/R ratios at both high and low metallicity (see \cite{Meynet:2011} and references therein).  In addition to metallicity, the B/R ratio is also dependent on mass and thus the star formation history (SFH) \citep{Langer:1995, Dohm-Palmer:2002, Vazquez:2007, McQuinn:2011}.  To compare the B/R ratios between galaxies, or even different regions in a single galaxy, one must be certain that the underlying SFH is identical.  Any systematic variation in the star formation history necessarily affects the radial dependence of the B/R ratio.  

\begin{figure}[ht!] 
\centering 
\subfigure{\includegraphics[width=0.49\columnwidth]{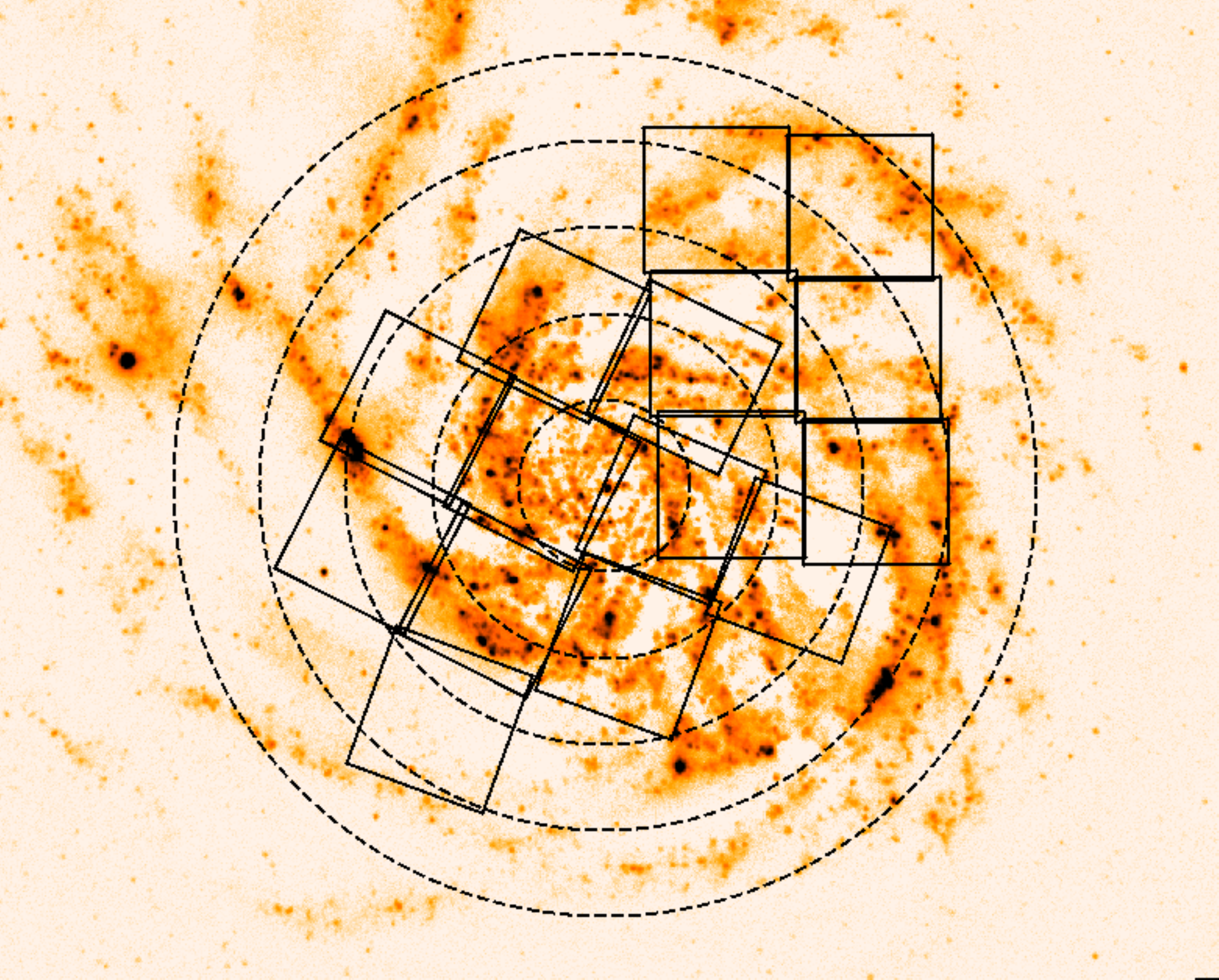}}
\subfigure{\includegraphics[width=0.49\columnwidth]{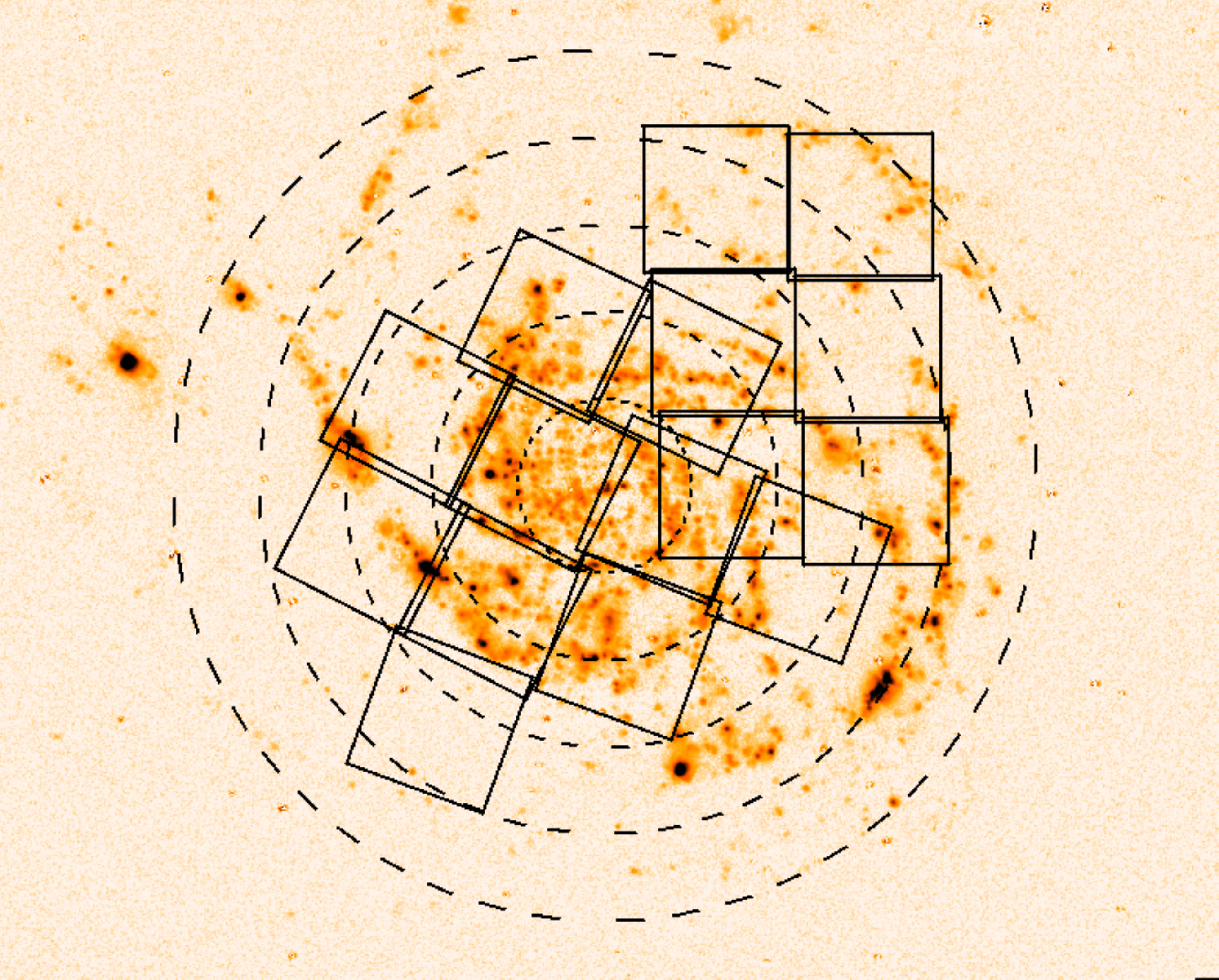}}

\caption{Images of M101: \textit{GALEX} FUV (left) and KPNO Schmidt H${\alpha}$ (right).  Overlaid are five dashed annuli 2$\arcmin$ wide and extend out to 10$\arcmin$.  The ACS fields-of-view are given by the solid black lines.  Pixel values are displayed logarithmically and range from zero to the maximum.  The contrast for both images is identical.} 
\label{fig:finderchart} 
\end{figure} 

Broad and narrow band imaging from IR to X-Ray suggest that similar to the stellar populations, the emission properties of galaxies vary radially.  Recently, deep observations of galaxies using \textit{Galaxy Evolution Explorer (GALEX)} have demonstrated ultraviolet (UV) emission that is considerably more extended than optical emission, a property characterized as an extended UV disk (XUV) \citep{Thilker:2005,Thilker:2007}.  Moreover, the UV disk is roughly twice that of H$\alpha$ \citep{Boissier:2007, Goddard:2010} suggesting that in the outskirts of XUV galaxies, O stars are rare but B and A type stars are abundant.  A truncated initial mass function (IMF) in low density star forming regions or statistical sampling of the IMF could suppress the formation of massive O stars, thereby leading to the absence of H$\alpha$ at large radii \citep{Boissier:2007, Meurer:2009, Goddard:2010, Lee:2011, Koda:2012}.  An alternative explanation is that the star forming regions are, on average, older in the outskirts \citep{Thilker:2005, Zaritsky:2007}.  Comparing the stellar populations in complexes observed in both UV and H$\alpha$ to complexes seen only in UV, suggests the lack of H$\alpha$ may be attributed to differences in stellar population age \citep{Thilker:2005, Gogarten:2009, Alberts:2011}.  Thus, discrepancies between the UV and H$\alpha$ profiles, and the radial dependence in the B/R ratio, may share a common origin: radial variations in the recent SFH.  

\begin{figure*}[ht!] 
\centering 
\subfigure{\includegraphics[width=0.18\textwidth]{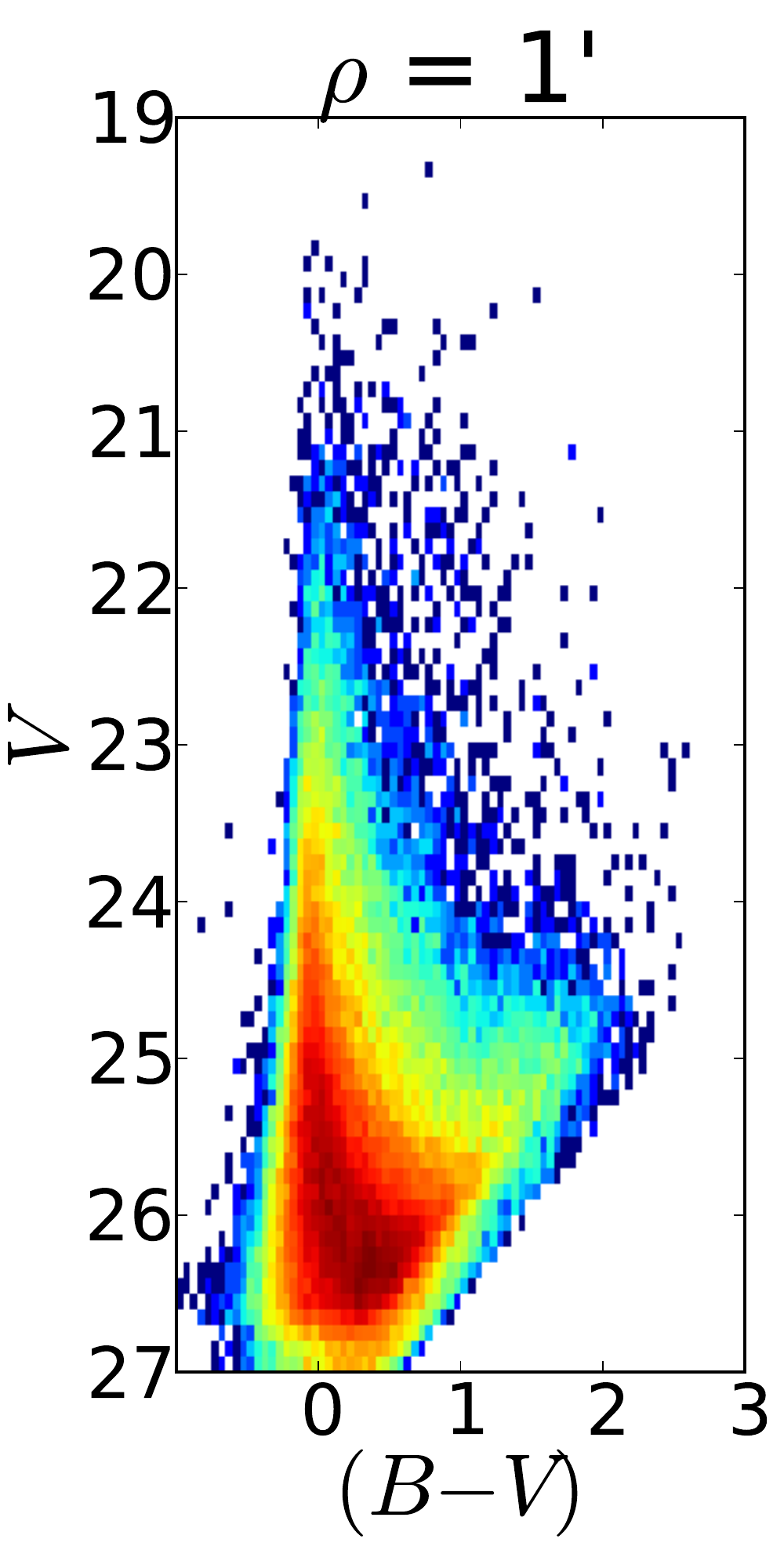}} 
\subfigure{\includegraphics[width=0.18\textwidth]{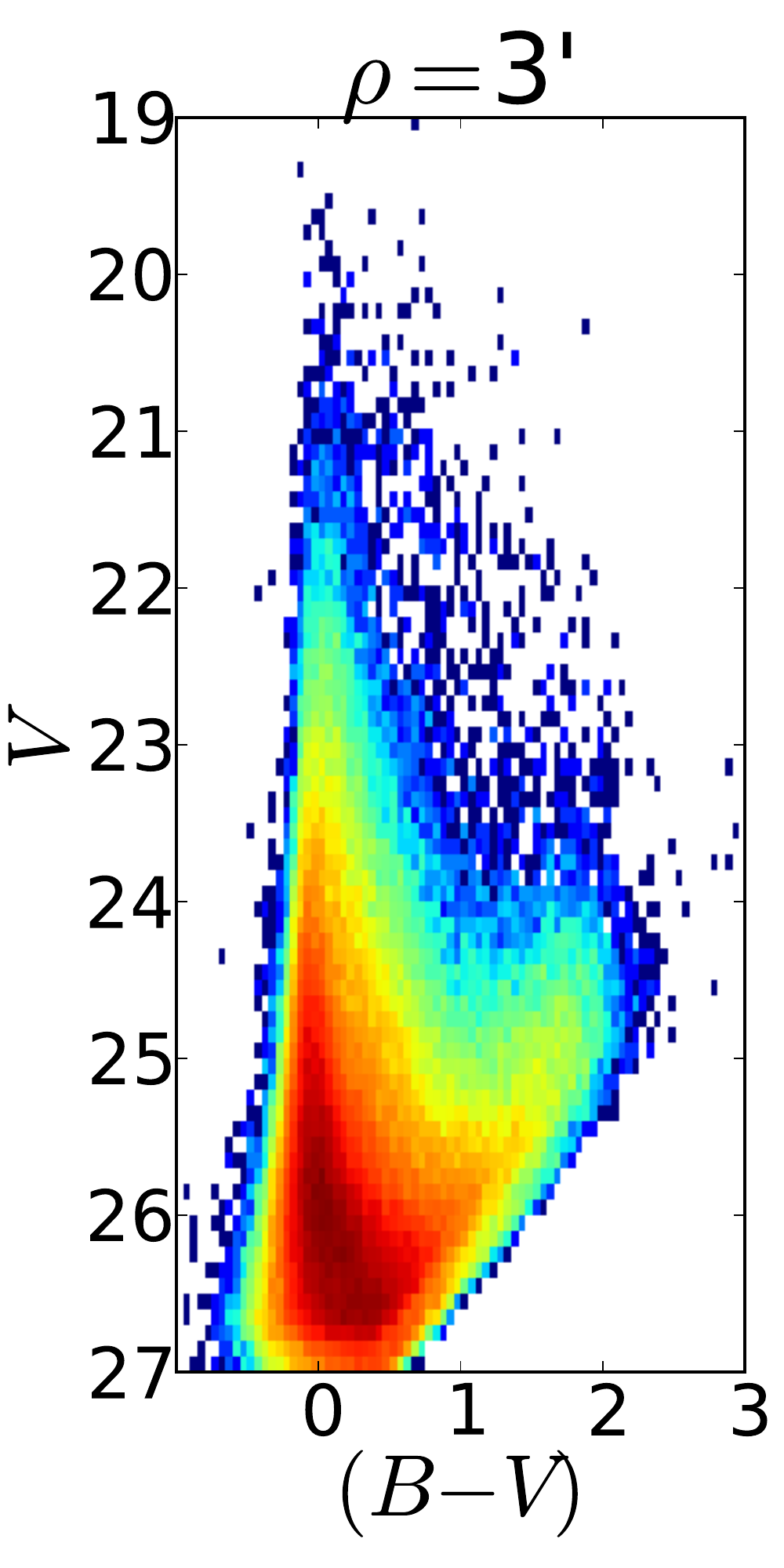}} 
\subfigure{\includegraphics[width=0.18\textwidth]{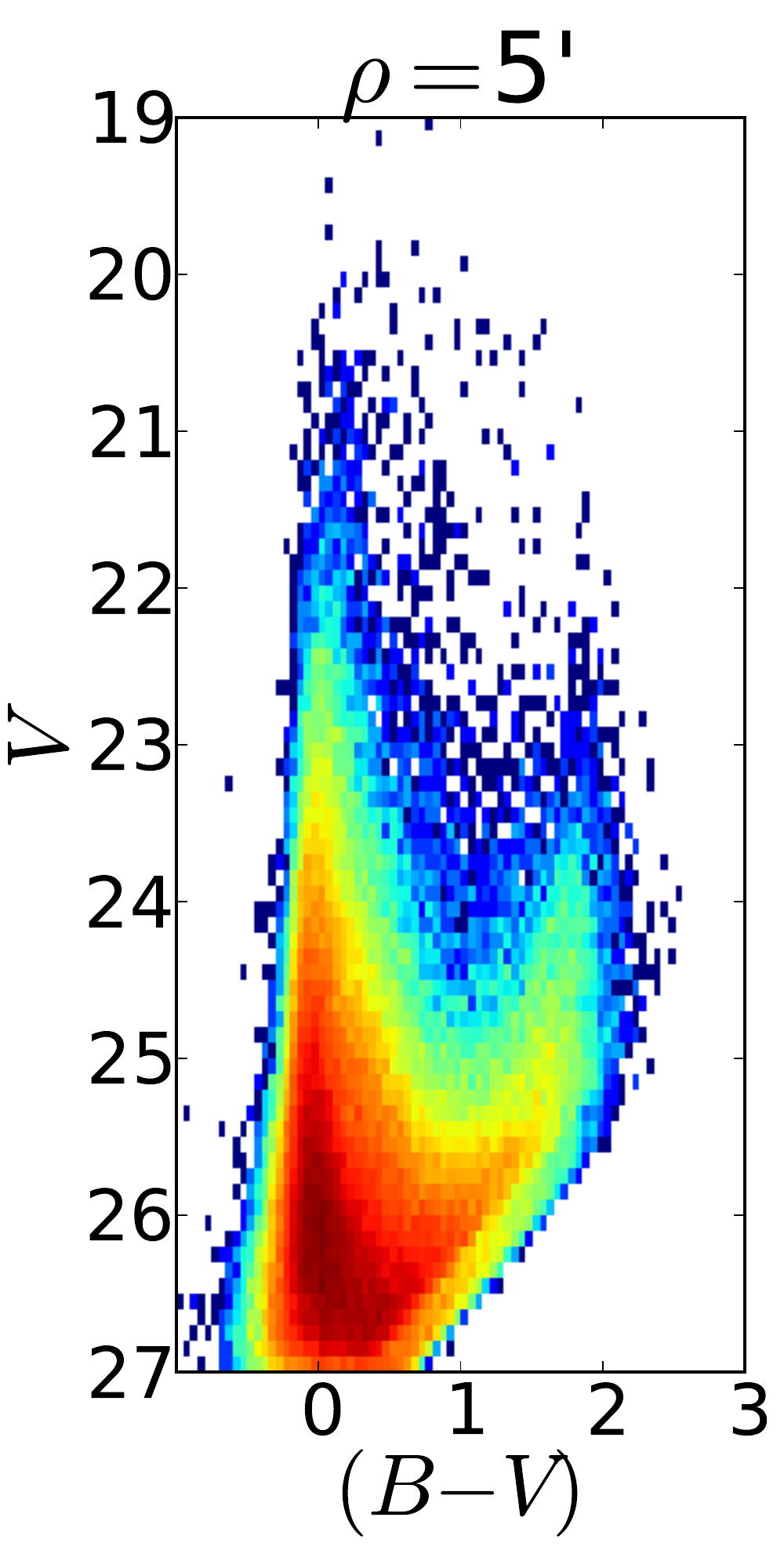}} 
\subfigure{\includegraphics[width=0.18\textwidth]{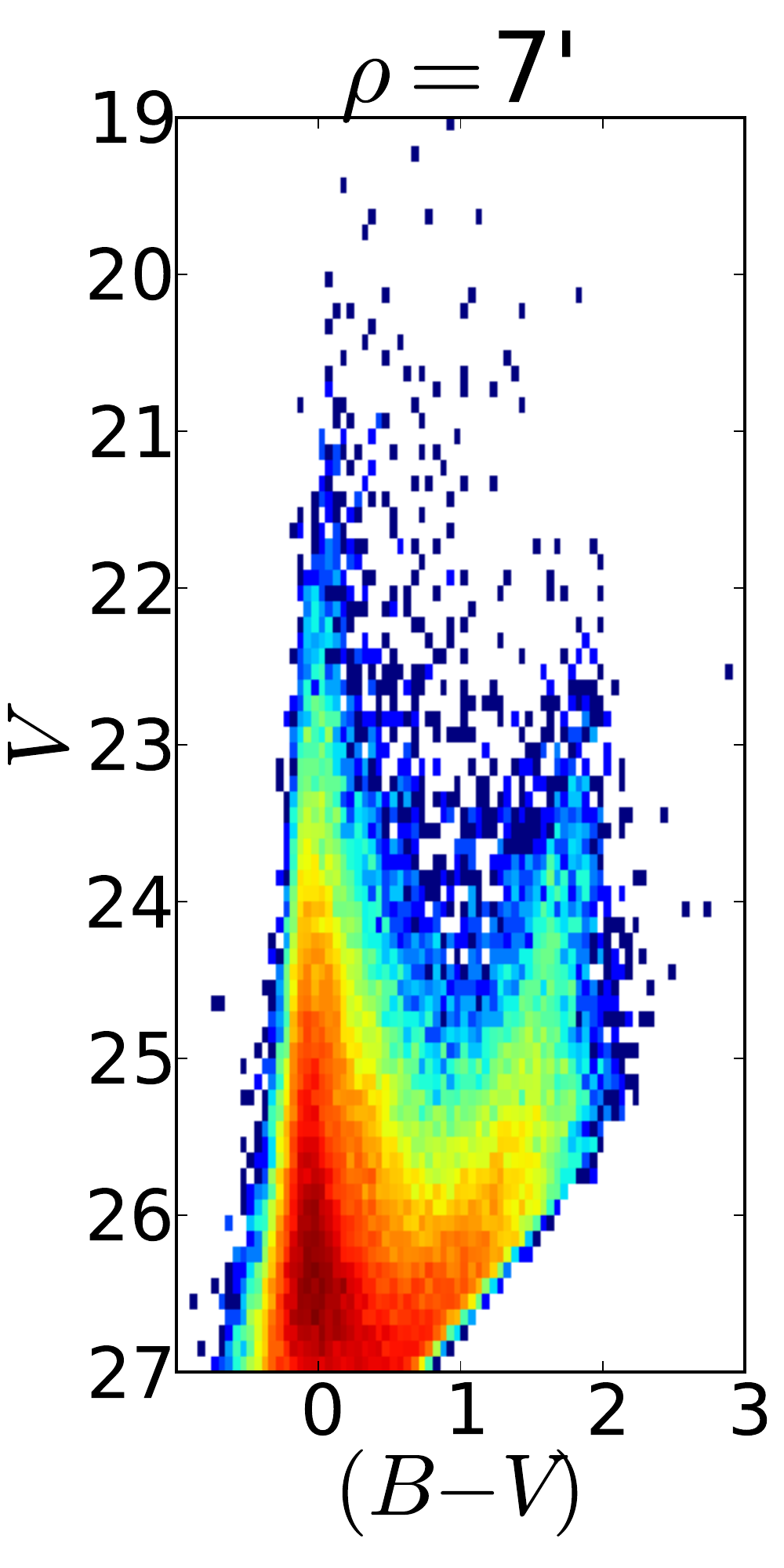}} 
\subfigure{\includegraphics[width=0.18\textwidth]{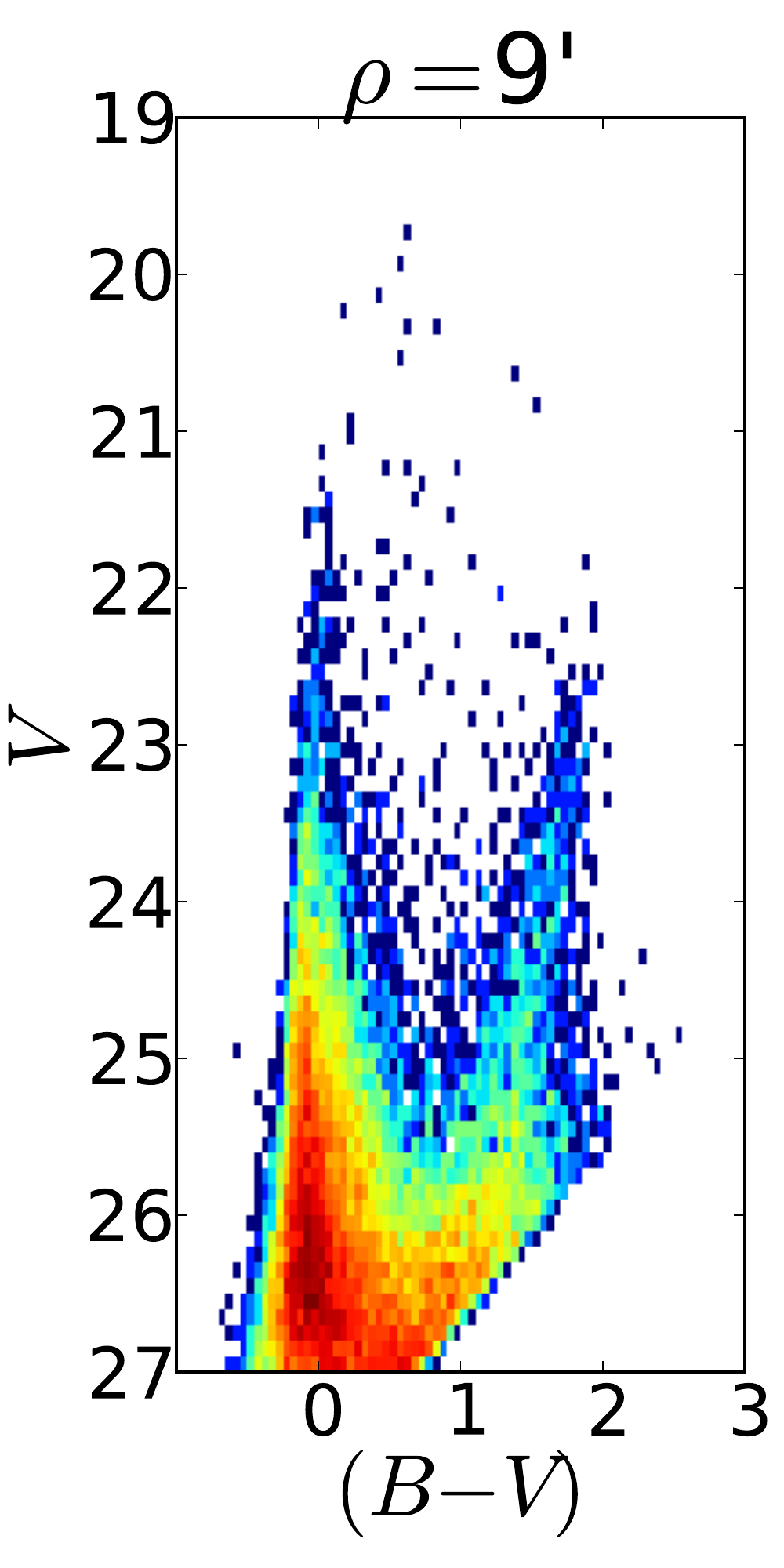}}

\caption{$V$ vs. $(B-V)$ CMDs for each annulus.  The radial center is denoted at the top of each CMD.  The color map is linear in stellar density with lighter colors indicating higher densities.  A color version is available online with blue indicating low stellar density and dark red indicating high stellar density.  The $V$ vs. $(V-I)$ CMDs are shown in the Appendix of the online version.} 
\label{fig:radial_Hess} 
\end{figure*} 
\begin{figure*}[ht!] 
\centering 

\subfigure{\includegraphics[width=0.18\textwidth]{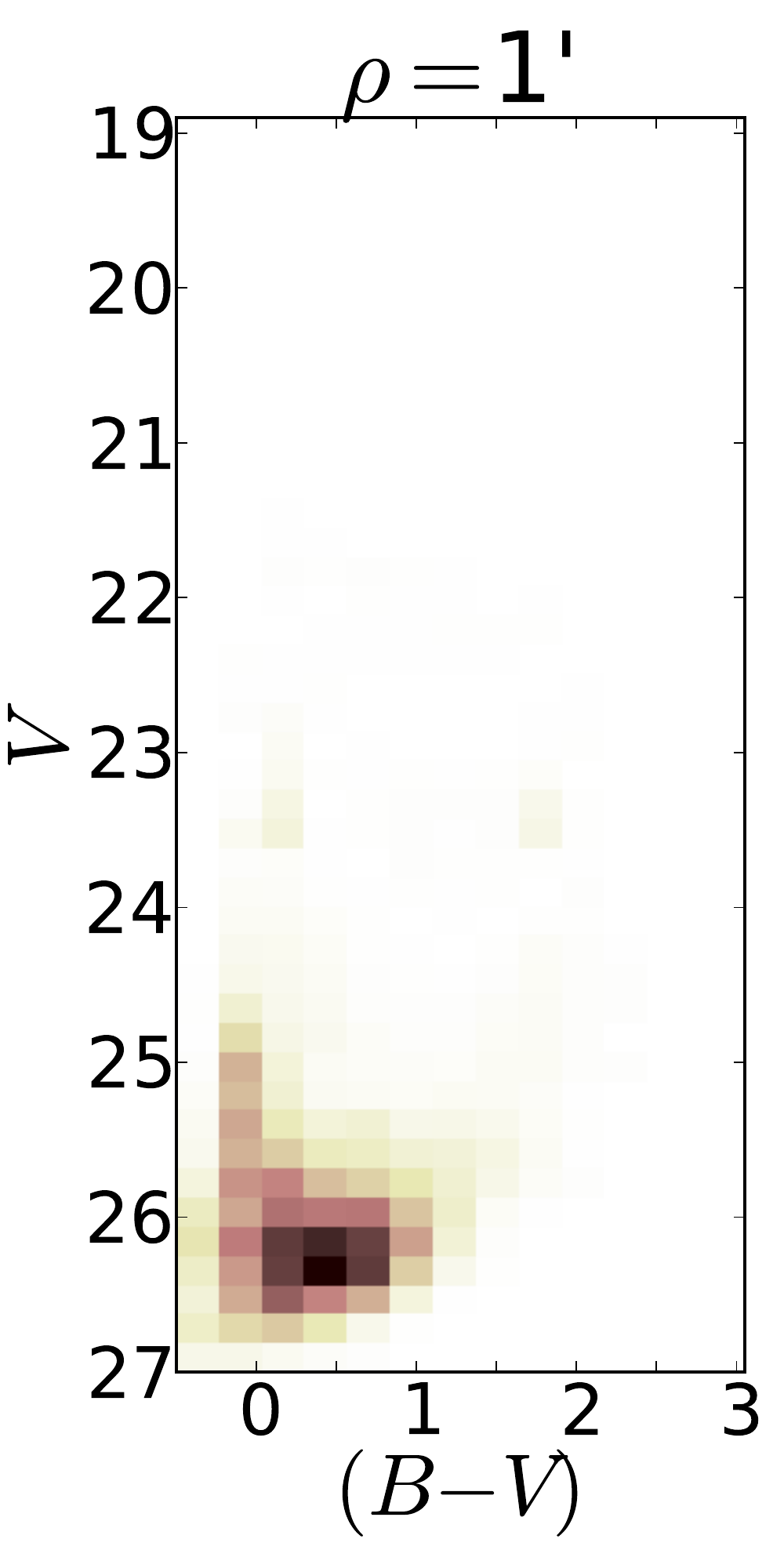}} 
\subfigure{\includegraphics[width=0.18\textwidth]{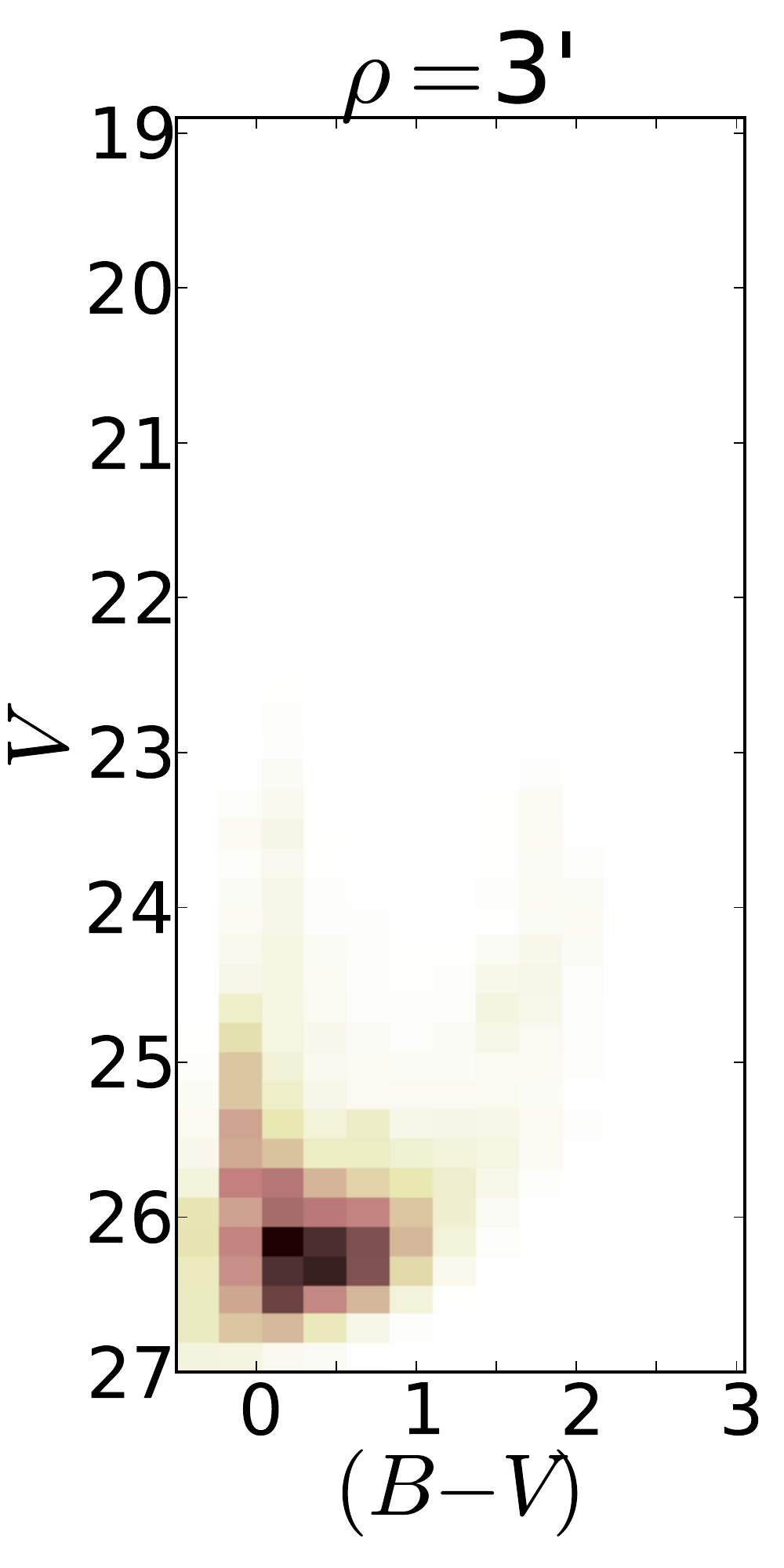}} 
\subfigure{\includegraphics[width=0.18\textwidth]{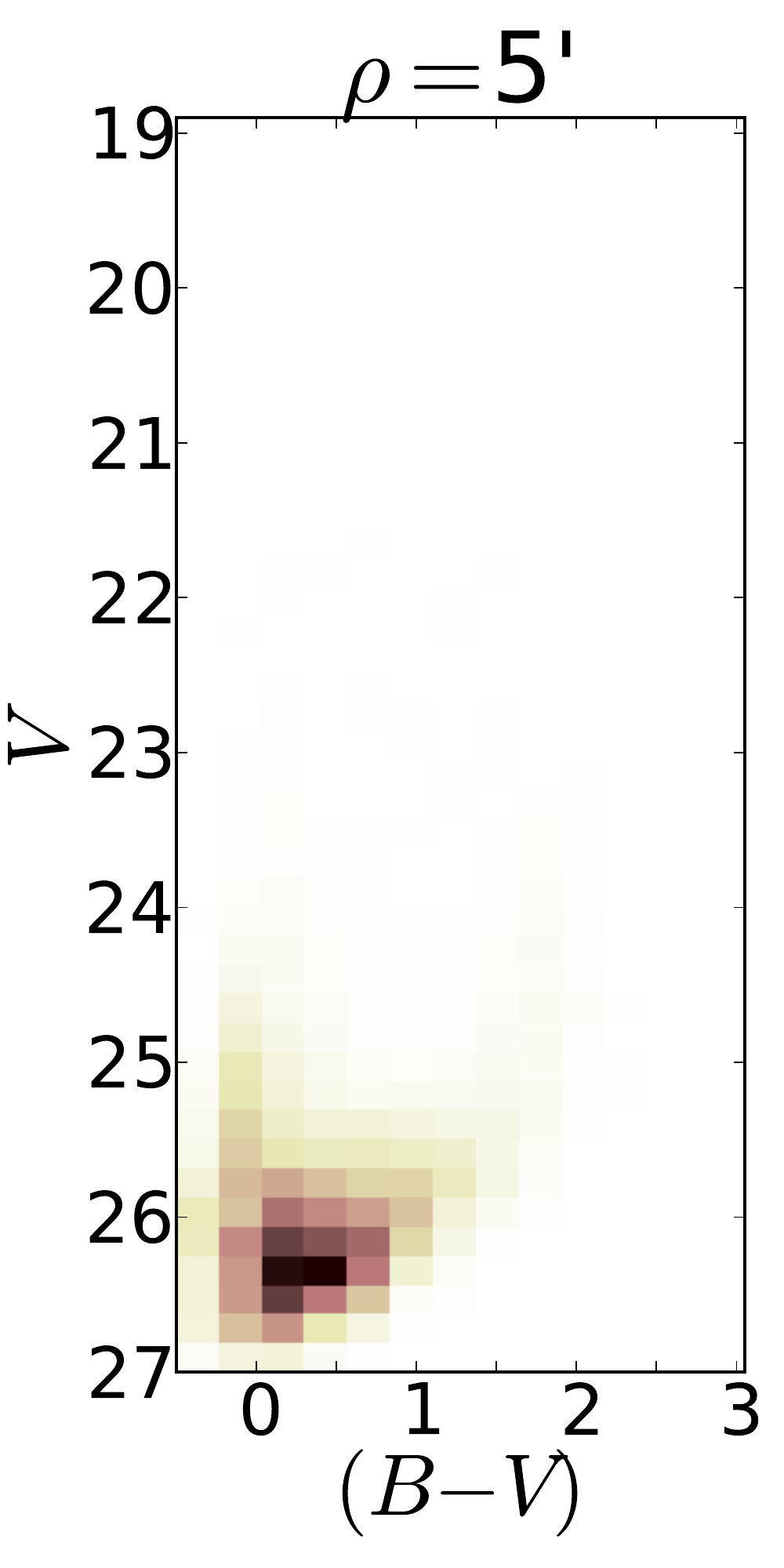}} 
\subfigure{\includegraphics[width=0.18\textwidth]{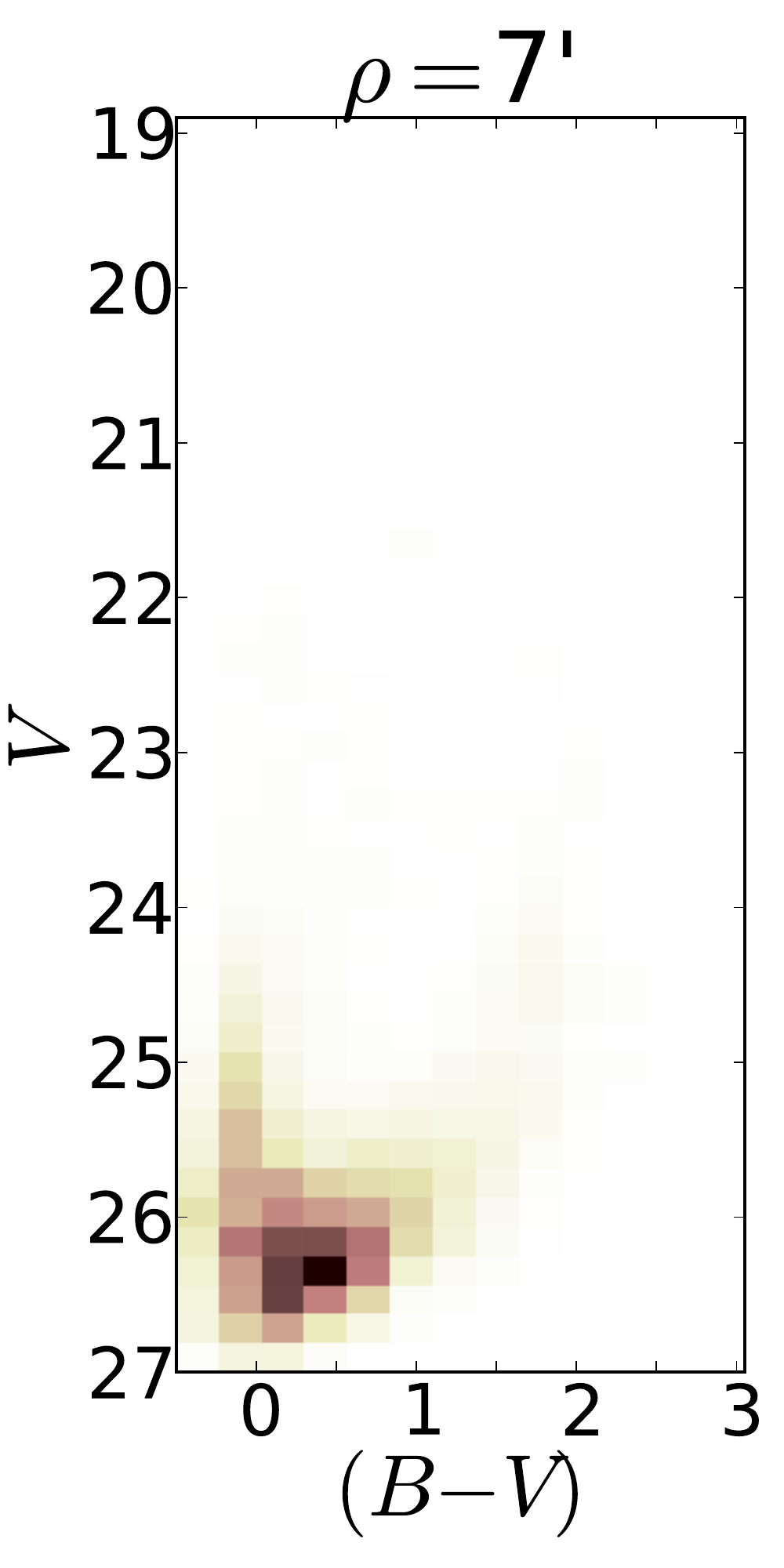}} 
\subfigure{\includegraphics[width=0.18\textwidth]{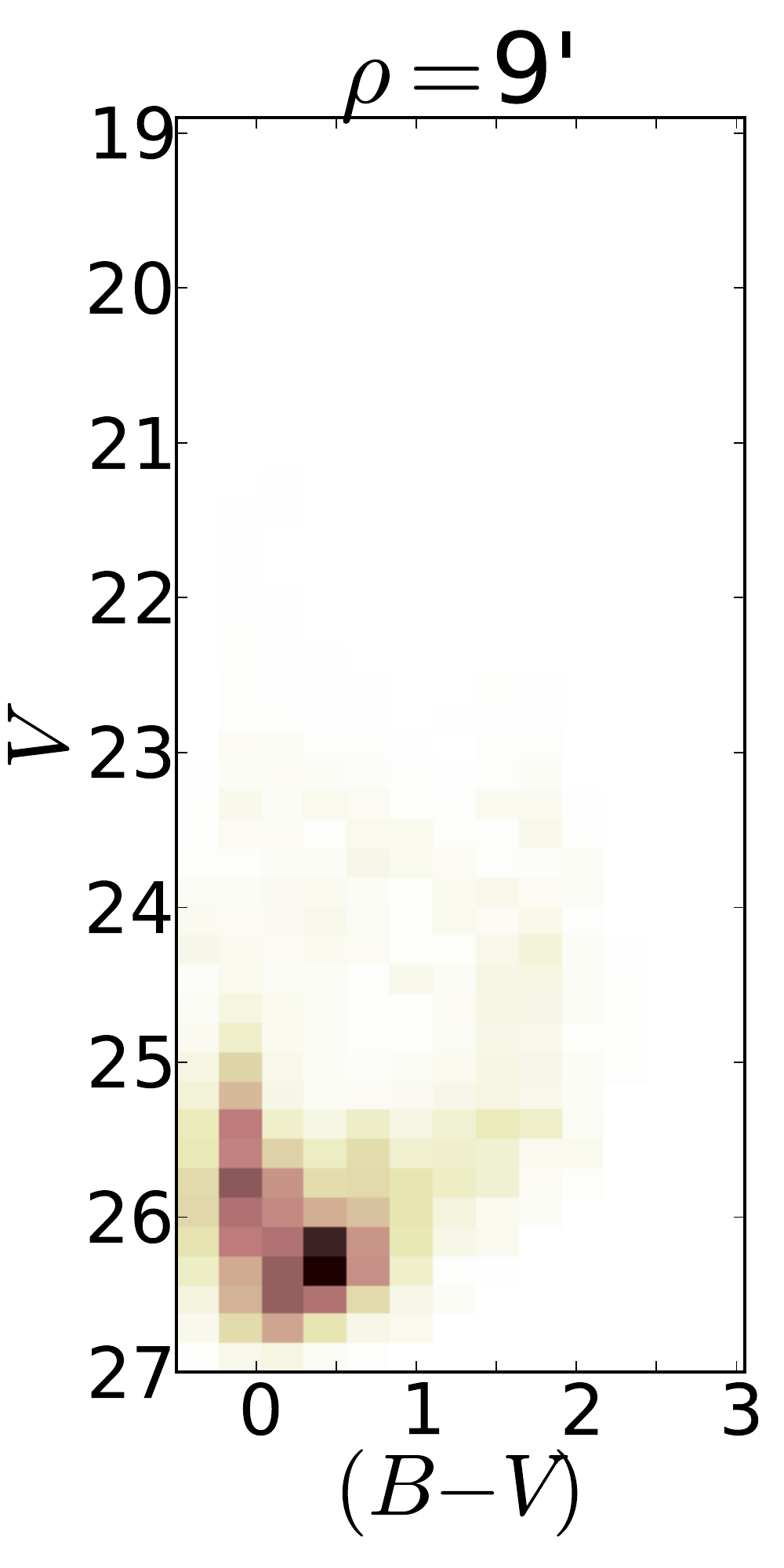}} 

\caption{Residual $V$ vs. $(B-V)$ CMDs.  Residuals were produced by taking the difference between the  model and observations and median filtering the results using a $0.25 \times 0.25$ mag$^{2}$ filter.  The color map is linear with darker regions indicating larger residuals.  The largest residuals systematically occur at magnitudes fainter than $V = 26$ due to incompleteness.} 
\label{fig:radial_res} 
\end{figure*} 

Multi-band photometry of M101 has demonstrated that not only does it possess an XUV disk \citep{Thilker:2007}, but that the H$\alpha$ emission is considerably less extended than the UV emission \citep{Mihos:2013} (Figure~\ref{fig:finderchart}).  Resolved stellar population studies have also shown that the B/R ratio decreases dramatically with radius \citep[hereafter Paper I]{Grammer:2013a}.  These properties make M101 an ideal target for studies of massive star formation and its connection between the observed stellar populations and global emission properties.  For this work, we aim to derive the radial SFH by modeling the resolved stellar populations in radial bins (annuli).  Our goal is to examine the variations in the radial SFH and their connection to the radial dependence in the B/R ratio as well as differences in the UV and H$\alpha$ radial profiles.  

The resolved stellar photometry is summarized in $\S2$.  In $\S3$ we discuss our methods for determining the SFH.  We examine the radial variations in the SFH in $\S4$. Finally, we evaluate the likelihood that the radial variations in the SFH are responsible for the differences in the H$\alpha$ and UV radial profiles as well as the shape of the B/R ratio and its dependence on radius.  The results are summarized in the last section.   

\begin{figure*}[ht!] 
\centering 
\subfigure{\includegraphics[width=0.35\textwidth]{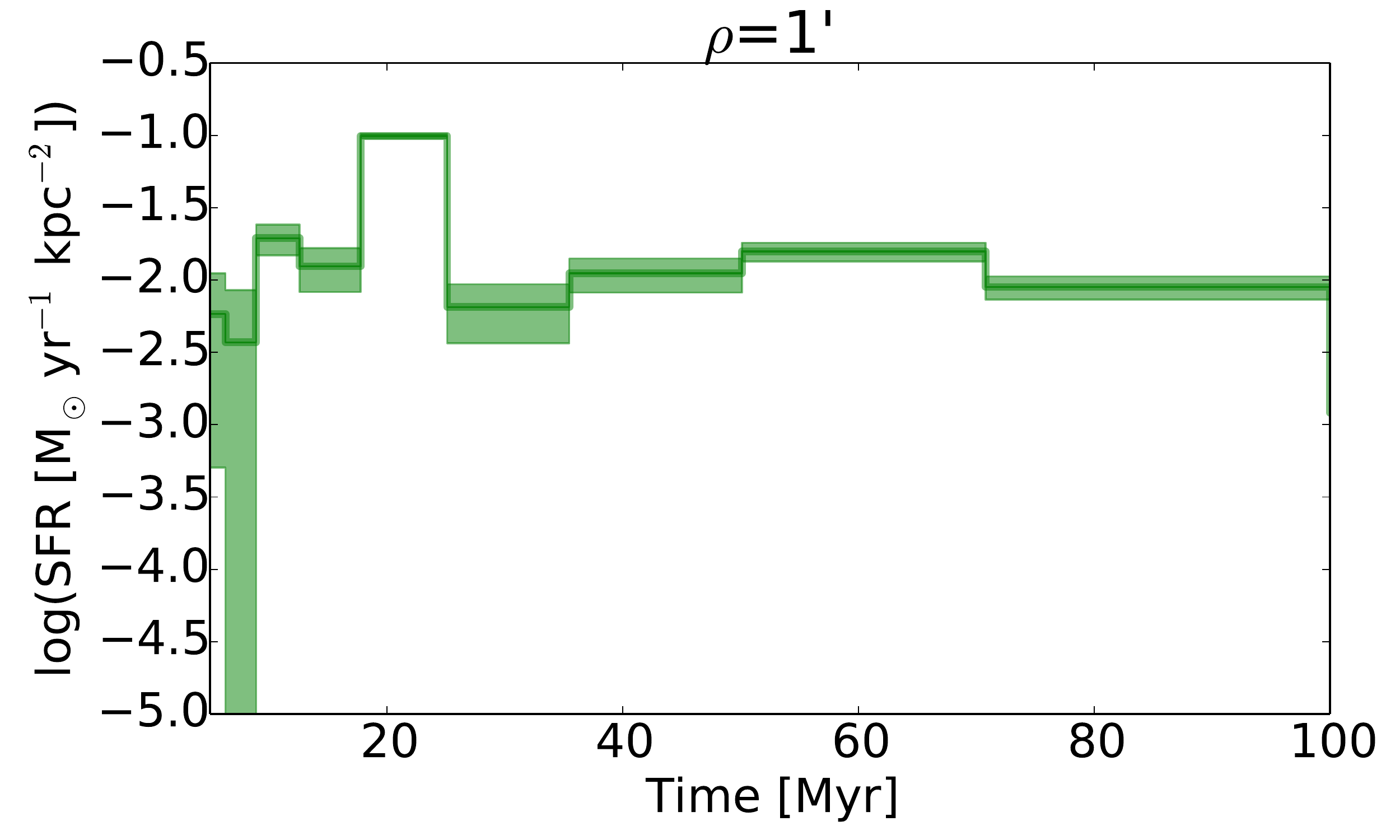}}\quad
\subfigure{\includegraphics[width=0.35\textwidth]{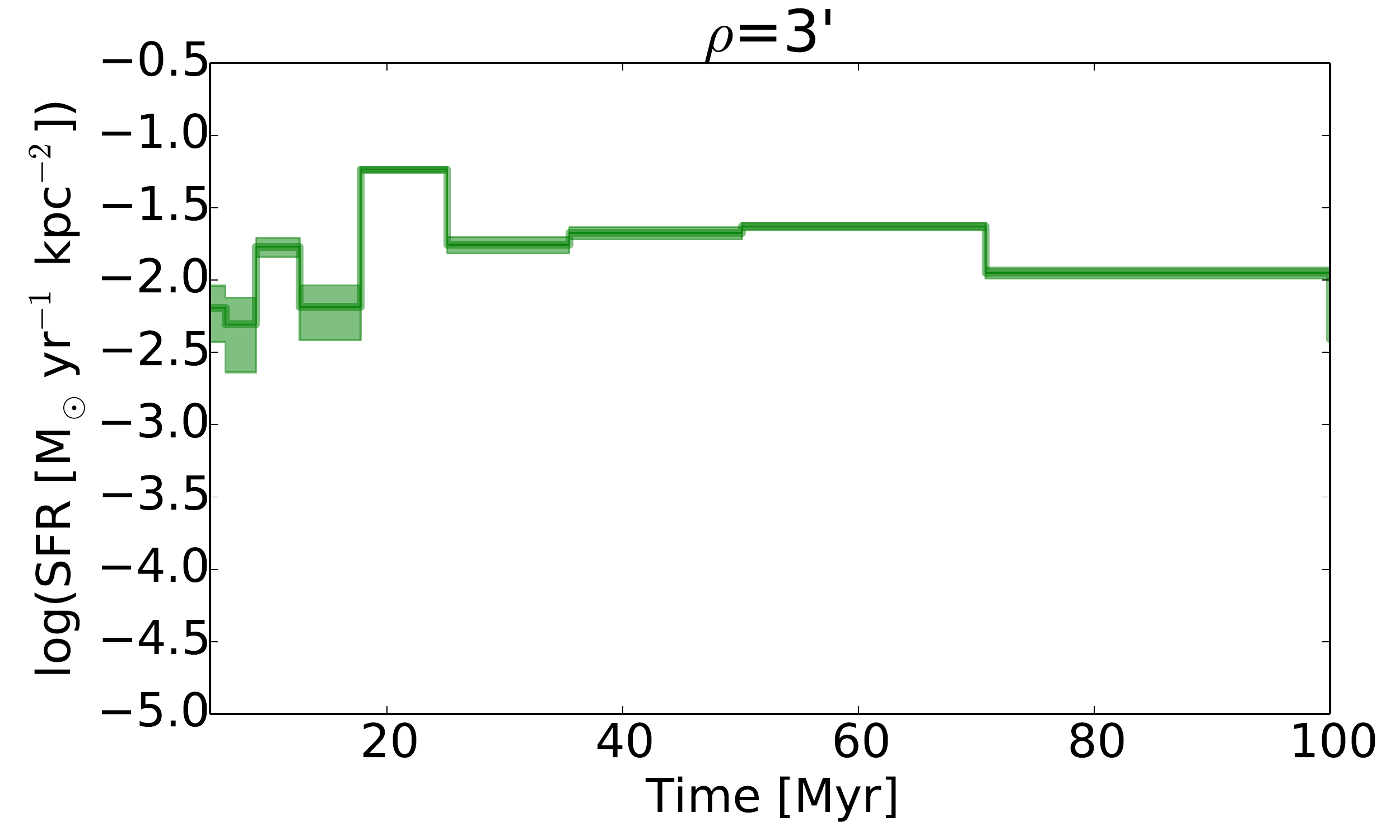}}\quad
\subfigure{\includegraphics[width=0.35\textwidth]{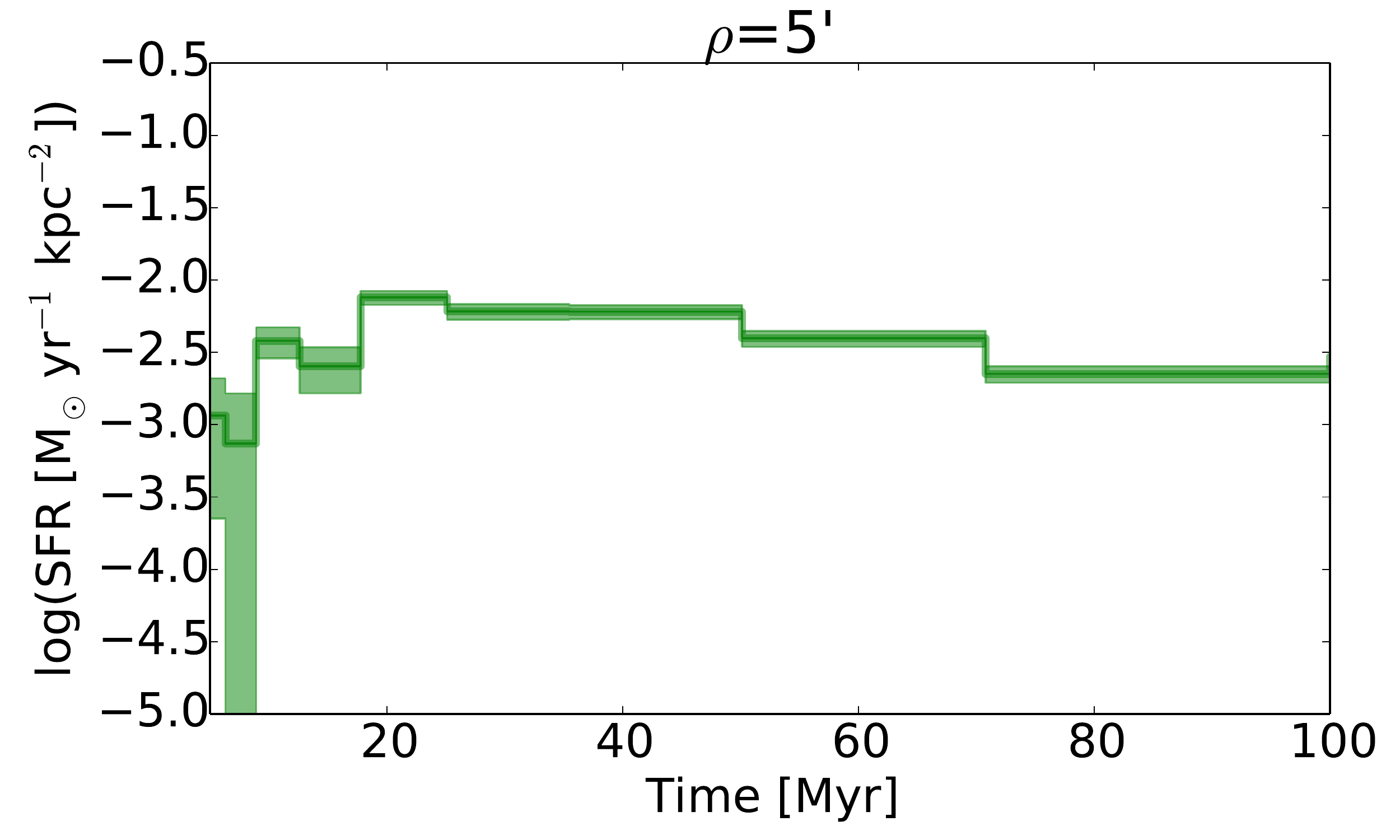}}\quad
\subfigure{\includegraphics[width=0.35\textwidth]{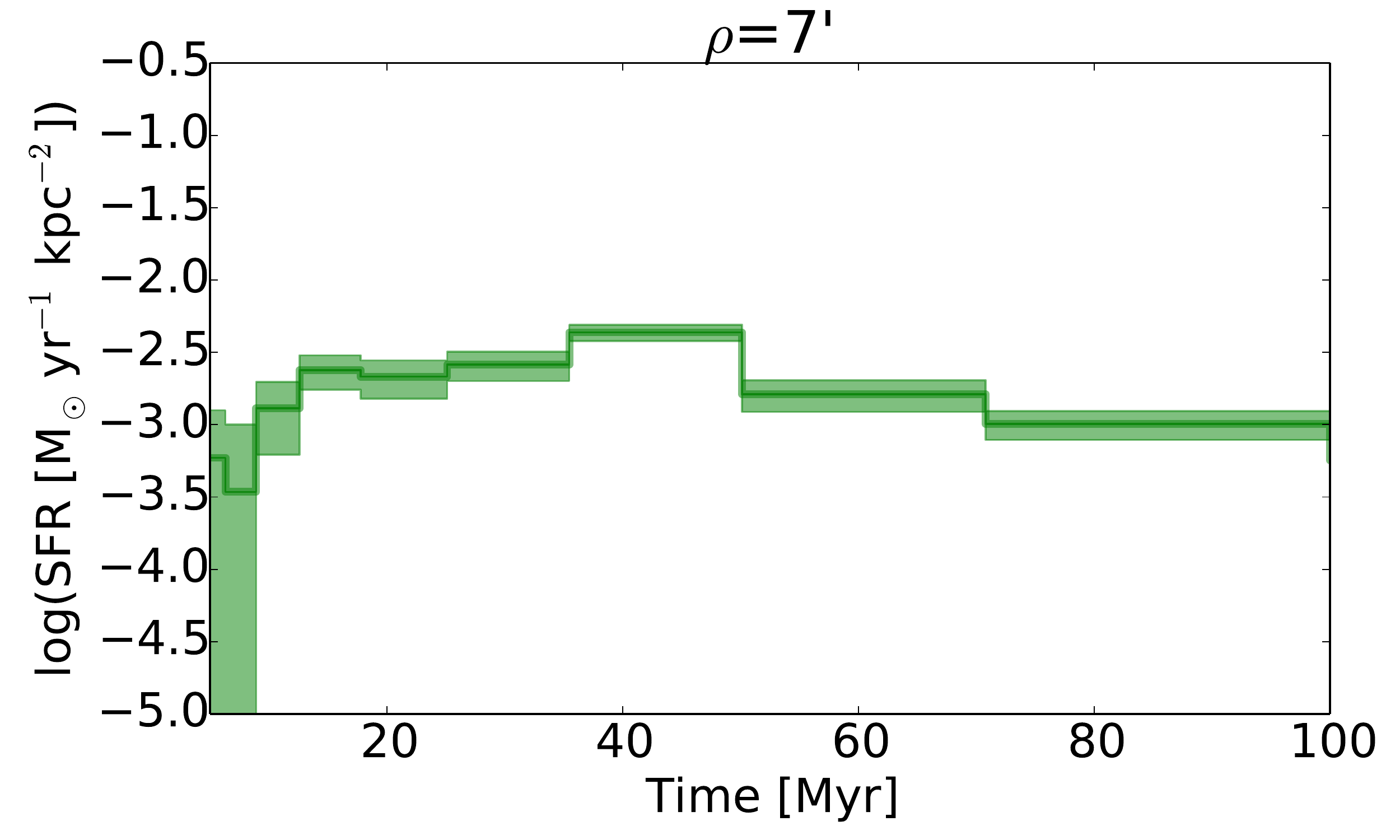}}\quad
\subfigure{\includegraphics[width=0.35\textwidth]{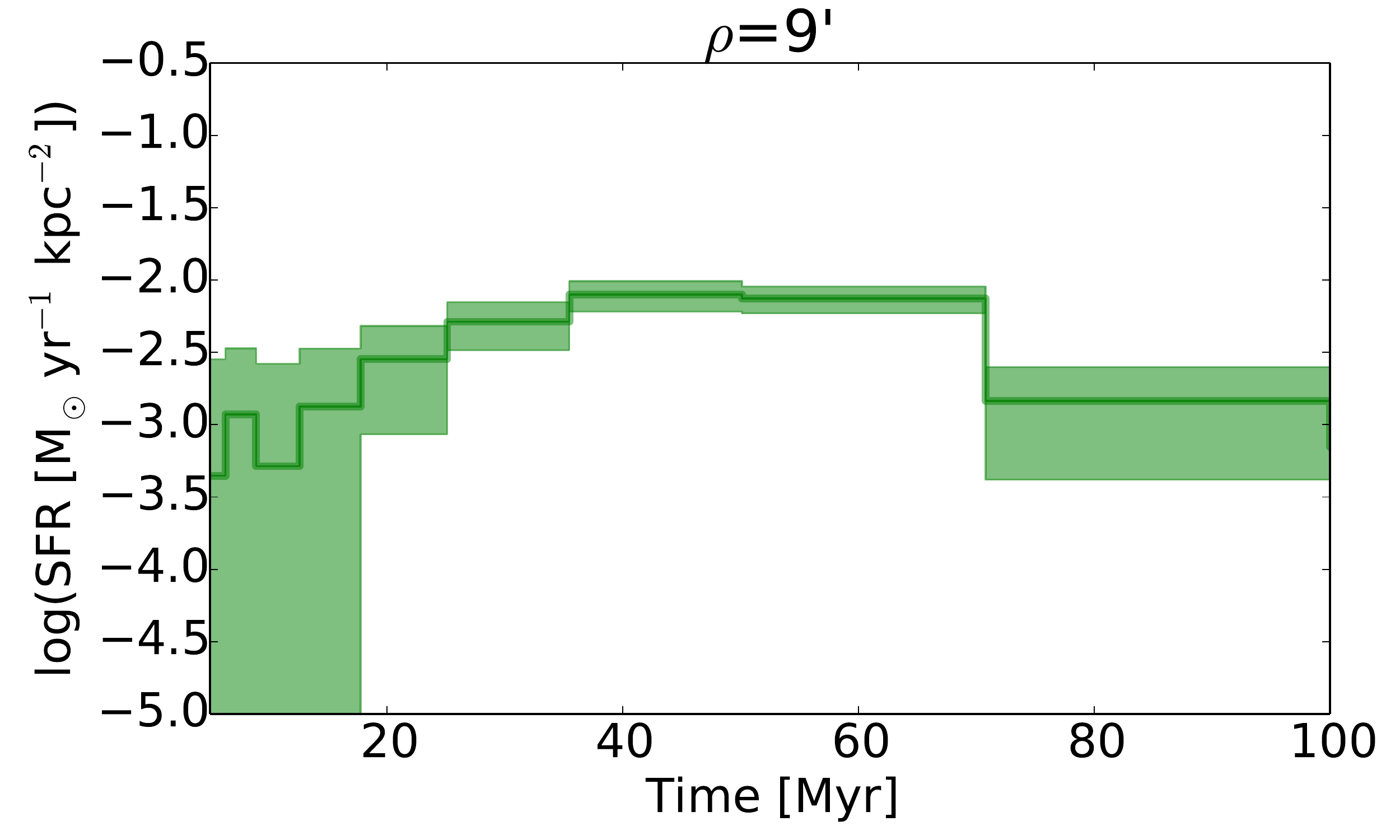}}\quad

\caption{The SFHs for the five annuli.  Each annulus is identified by its radial mid-point in arcminutes.  The SFHs were determined by comparing observed CMDs to synthetic CMDs.  The shaded regions denote the 68$\%$ confidence intervals.} 
\label{fig:radialSFH}
\end{figure*} 

\section{Data and Photometry}
We used archival observations of M101 from the \textit{The Hubble Space Telescope (HST)} Advanced Camera for Surveys (ACS) Wide Field Camera to produce a catalog of $\sim500000$ resolved stars \footnotemark[1].  Our catalog has photometric errors that are less than 10$\%$ for $B < 24.5$, $V < 24.3$, and $I < 24.0$ and is 50$\%$ complete down to $B = 27.0$, $V = 26.5$, and $I = 26.2$ for the fields from proposal ID 9490; fields from proposal ID 9492 are 50$\%$ complete down to $B = 27.3$, $V = 26.8$, and $I = 26.4$. See Paper I for details regarding our photometric technique.

\footnotetext[1]{http://etacar.umn.edu/LuminousStars/M101/M101-HST-ACS-WFC- Catalog.txt}

We assigned galactocentric angular distances to each star assuming an inclination angle of 18$^{\circ}$ and a position angle of 39$^{\circ}$ \citep{Bosma:1981}.  We then divided the stars into five $2\arcmin$ wide annuli (see Figure~\ref{fig:finderchart}).  At a distance to M101 of 6.5 Mpc \citep{Shappee:2011}, each annulus has a width of $\sim$1.7 kpc, and the five annuli span a radial distance of $\sim$17 kpc.  For each annulus, we calculate the observed area taking into account the coverage of the ACS fields.  We identify each annulus by its radial mid-point in arcminutes.  In Figure~\ref{fig:radial_Hess} we display the three-dimensional color-magnitude diagrams (CMDs) for each annulus: the color-coded dimension indicates the density of stars, in color-magnitude space, scaled by peak stellar density.  

\section{Modeling the Color-Magnitude Diagrams}
\subsection{Method}
The method of modeling observed CMDs with synthetic CMDs to extract the SFH is a well established technique \citep{Tosi:1991, Gallart:1999, Hernandez:1999, Holtzman:1999, Dolphin:2002, Skillman:2003, Harris:2004}.  We used StarFISH \citep{Harris:2004} to derive the SFH for each of our annuli.  The user-supplied parameters to StarFISH are age, stellar mass and metallicity range, the slope of the IMF, binary fraction, photometric error functions, extinction, and the distance modulus. Single-age synthetic CMDs are then created from the given parameters.  The observed CMDs are modeled as the weighted sum of all the synthetic CMDs while simultaneously minimizing $\chi^{2}$.  The fit with the lowest $\chi^{2}$ is returned as the most likely SFH.

Though we are most concerned with the massive star population of M101, our CMDs contain stellar populations with a range of ages and masses.  Thus, to fit the CMDs we assume an IMF slope of -2.35 \citep{Salpeter:1955} for stellar masses $0.1-100 $M$_{\odot}$, a binary fraction of 0.5, and a distance modulus of $\mu_{0} = 29.05$ \citep{Shappee:2011}.  Crowding, completeness, and observational errors are incorporated into the model CMDs by supplying StarFISH with the input and output magnitudes of our artificial star tests as well as whether the artificial stars passed our quality criteria (discussed in Paper I).  From these parameters, CMDs were synthesized using the Padova stellar isochrones \citep{Marigo:2008} with metallicities in the range of $0.005 - 0.020$, and ages $4-500$ Myr spaced logarithmically.  We eliminate metallicity as a free parameter by assigning values from the observed abundance gradient \citep{Kennicutt:2003} to each annulus.  Similarly, we account for extinction by applying the radial extinction curve from \cite{Lin:2013} to the library of synthetic CMDs.  The library of synthetic CMDs was then used to model the stellar populations in each annulus.  The output of StarFISH is the amplitude for each time bin, in numbers of stars, which was converted to SFR by multiplying by the mean mass and dividing by the width of the time bin.  The uncertainties associated with each amplitude is the $68\%$ confidence interval resulting from Monte Carlo simulations performed by StarFISH.

\subsection{Evaluating the Reliability of the Star Formation Histories}
We assess how well the model CMDs reproduce the observations by simulating observations drawn from the best-fit SFH.  We then created residuals diagrams by subtracting the observed CMDs from the model CMDs, scaled by the number of observed stars in each color-magnitude bin.  The normalized residuals are displayed in Figure~\ref{fig:radial_res} after having applied a $0.25 \times 0.25$ mag$^{2}$ median filter to increase the contrast.  Based on Figure~\ref{fig:radial_res}, the model CMDs adequately reproduce the observations for magnitudes brighter than $V\approx26$, where residuals are small ($<10\%$).  At $V\approx26$, roughly corresponding to the main-sequence turn off of the 100 Myr isochrone, the residuals increase rapidly, which we attribute to incompleteness (see $\S2$).
 
\begin{figure}[ht!] 
\centering 
\subfigure{\includegraphics[width=0.75\columnwidth]{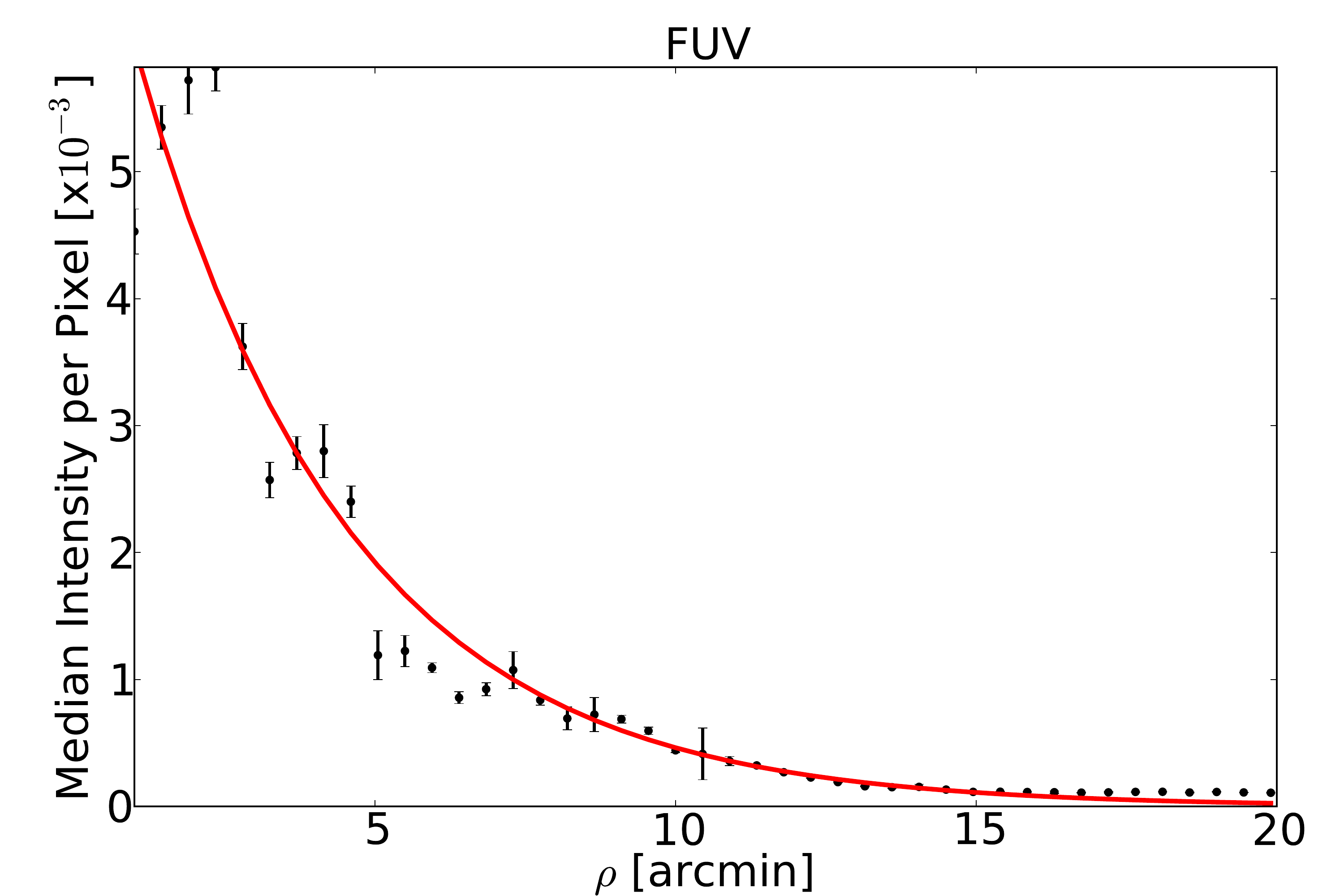}}
\subfigure{\includegraphics[width=0.75\columnwidth]{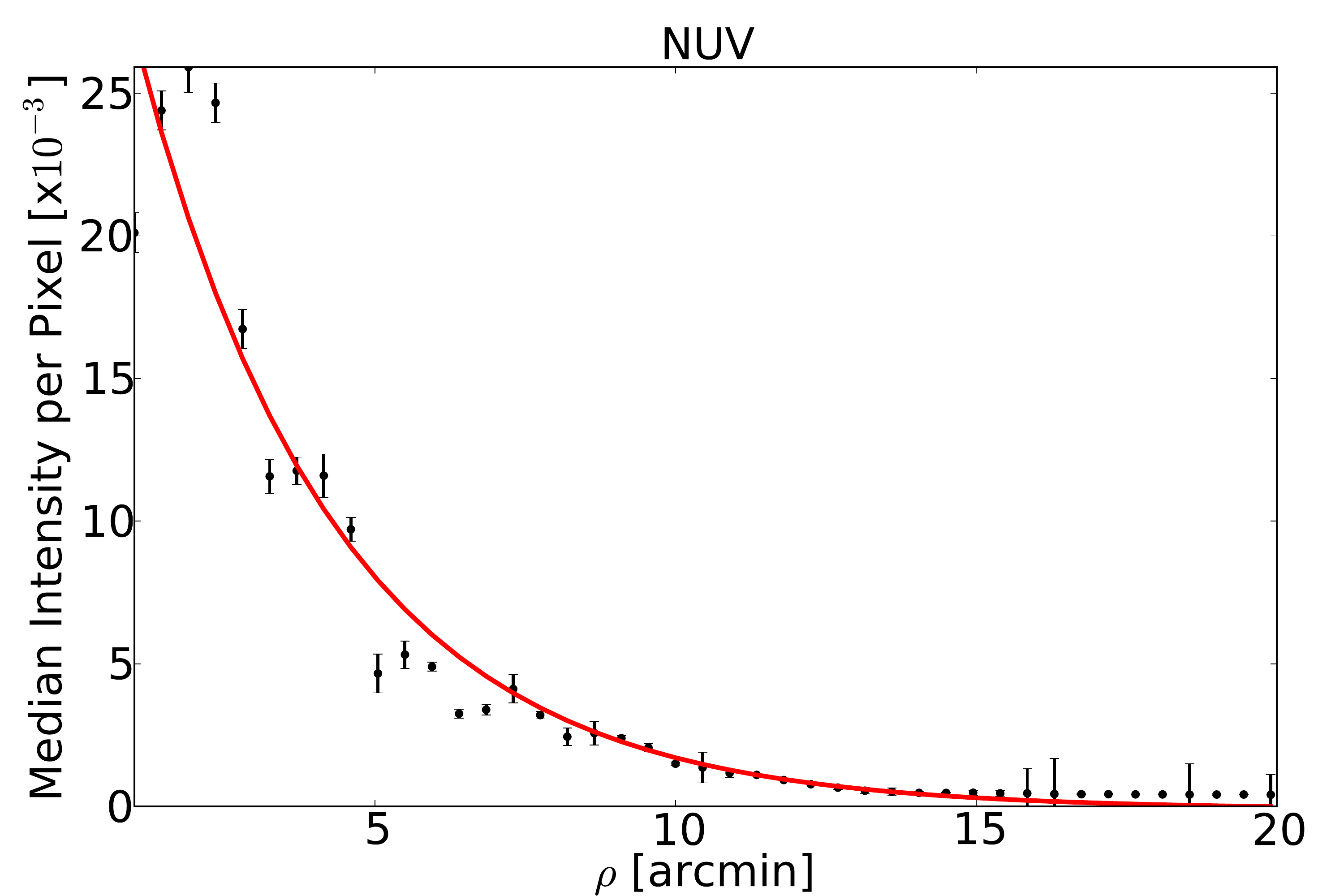}}
\subfigure{\includegraphics[width=0.75\columnwidth]{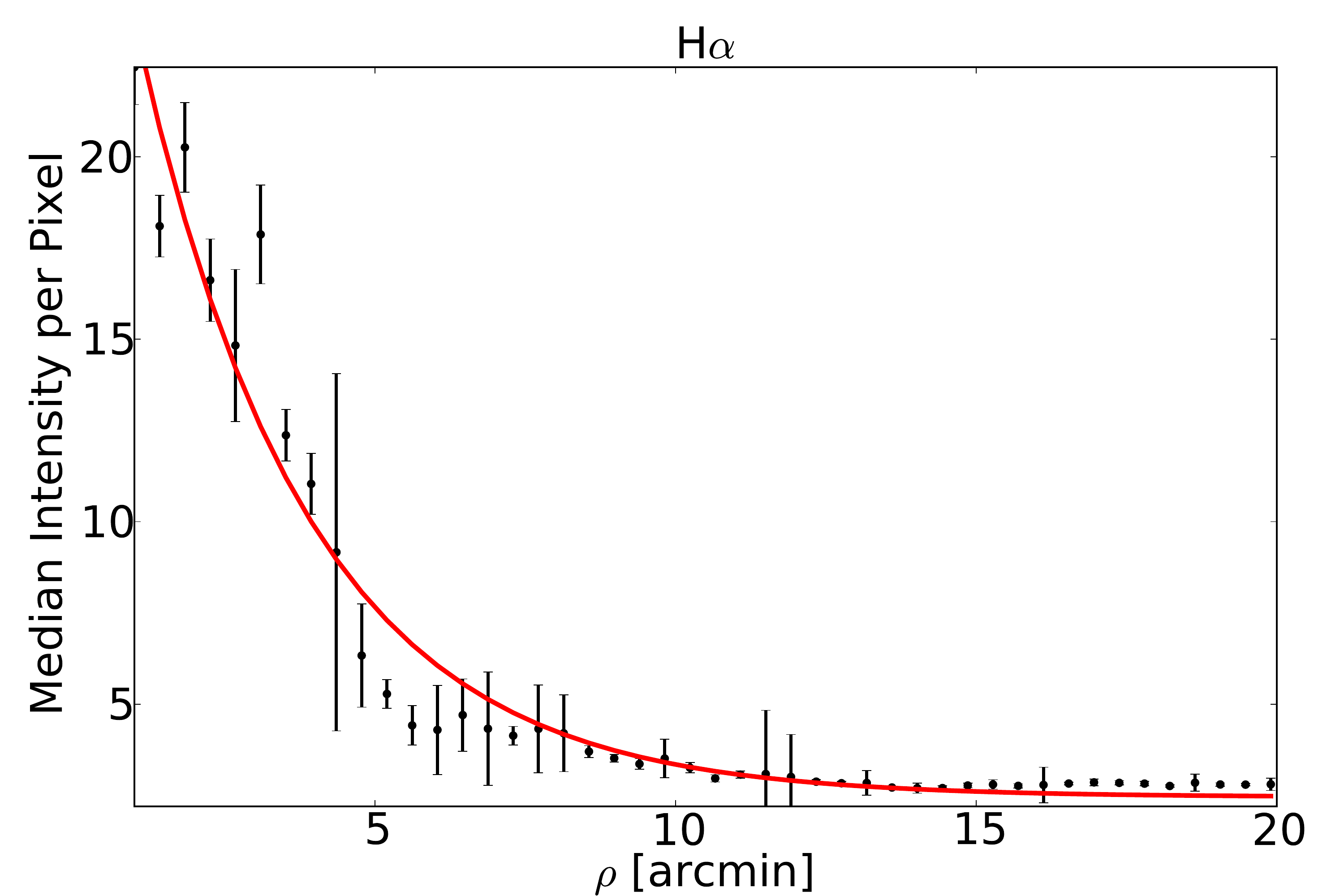}}

\caption{Radial profiles for FUV, NUV, and H${\alpha}$.  The solid lines are the best-fit exponential profiles.  The best-fit scale lengths, in arcminutes, are $3.36\pm0.18$, $3.51\pm0.23$, and $2.62\pm0.32$, respectively.  At the assumed distance of M101, the scale lengths correspond to $6.32\pm0.34$, $6.60\pm0.43$, and $4.93\pm0.60$ kpc.} 
\label{fig:radial_profiles}
\end{figure} 

The reliability of the reported SFRs were evaluated by performing a series of Monte Carlo simulations.  We created synthetic CMDs, assuming a constant SFR, which we then used as input to StarFISH.  We then compared the input and output SFH.  This test was performed one hundred times for SFRs between $10^{-2} - 10^{-7}$ M$_{\odot}$ yr$^{-1}$ in steps of 0.5 dex.  Our results indicate that SFRs higher than $\sim10^{-4.5}$ M$_{\odot}$ yr$^{-1}$ were consistently recovered with errors less than $15\%$ for all ages and metallicities.  Lower SFRs frequently had recovered values that differed from the input by up to $50\%$.  Given the range in observed surface areas, the SFR of our annuli at $1\arcmin$ and $9\arcmin$ are reliable when greater than $\sim10^{-6}$ M$_{\odot}$ yr$^{-1}$ kpc$^{-2}$.  The remaining annuli ($3\arcmin$, $5\arcmin$, and $7\arcmin$) have SFRs that are reliable down to $\sim10^{-7}$ M$_{\odot}$ yr$^{-1}$ kpc$^{-2}$.

\section{Results}
\subsection{The Star Formation Histories} 
Since the residuals in our CMD fits are low for magnitudes brighter than $V \approx 26$, and stellar populations older than 100 Myr do not contribute to the UV and H${\alpha}$ emission, we only display the last 100 Myr (Figure~\ref{fig:radialSFH}).  The overall behavior of the SFH has a clear radial dependence.  The two inner annuli display discrete bursts of star formation with peaks in SFR around 10 Myr ago and again at 20 Myr ago.  At $\rho = 5\arcmin$, the peaks at 10 Myr and 20 Myr are visible in the SFH but are considerably less pronounced than those observed in the inner radii.  In the outer annuli, the change in SFR is more gradual with time, although we find that the annulus at $\rho = 7\arcmin$ displays a burst in SFR at 40 Myr.  Similarly, the $\rho = 9\arcmin$ annulus shows a sudden increase in SFR starting at 60 Myr ago, followed by a gradual decrease in SFR at more recent times.  If we compare the SFR $<35$ Myr ago to SFR $35-100$ Myr ago, we see that there is a decline in the relative proportion of recent star formation with radius.  Consequently, a radial age gradient may exist for stars $<100$ Myr old.  In the following sections, we perform a more in-depth inspection of the SFH for evidence of a radial age gradient.

\begin{deluxetable}{lc|cc}
\centering
\tablewidth{0pt}
\tabletypesize{\normal}
\tablecolumns{4}
\tablecaption{Best-Fit Exponential Scale Lengths \label{tab:scalelengths}}
\tablehead{\colhead{Time Frame} & \colhead{Scale Length (kpc)} & \colhead{Filter} & \colhead{Scale Length (kpc)}}

\startdata
  $<10$ Myr  & $4.65\pm0.21$ & H$\alpha$   & $4.93\pm0.60$ \\ 
  $<35$ Myr  & $6.38\pm0.26$ & FUV         & $6.32\pm0.34$ \\
  $<100$ Myr & $6.30\pm0.15$ & NUV         & $6.63\pm0.43$
\enddata
\end{deluxetable}
\begin{figure}[htb!]
\centering

\includegraphics[width=\columnwidth]{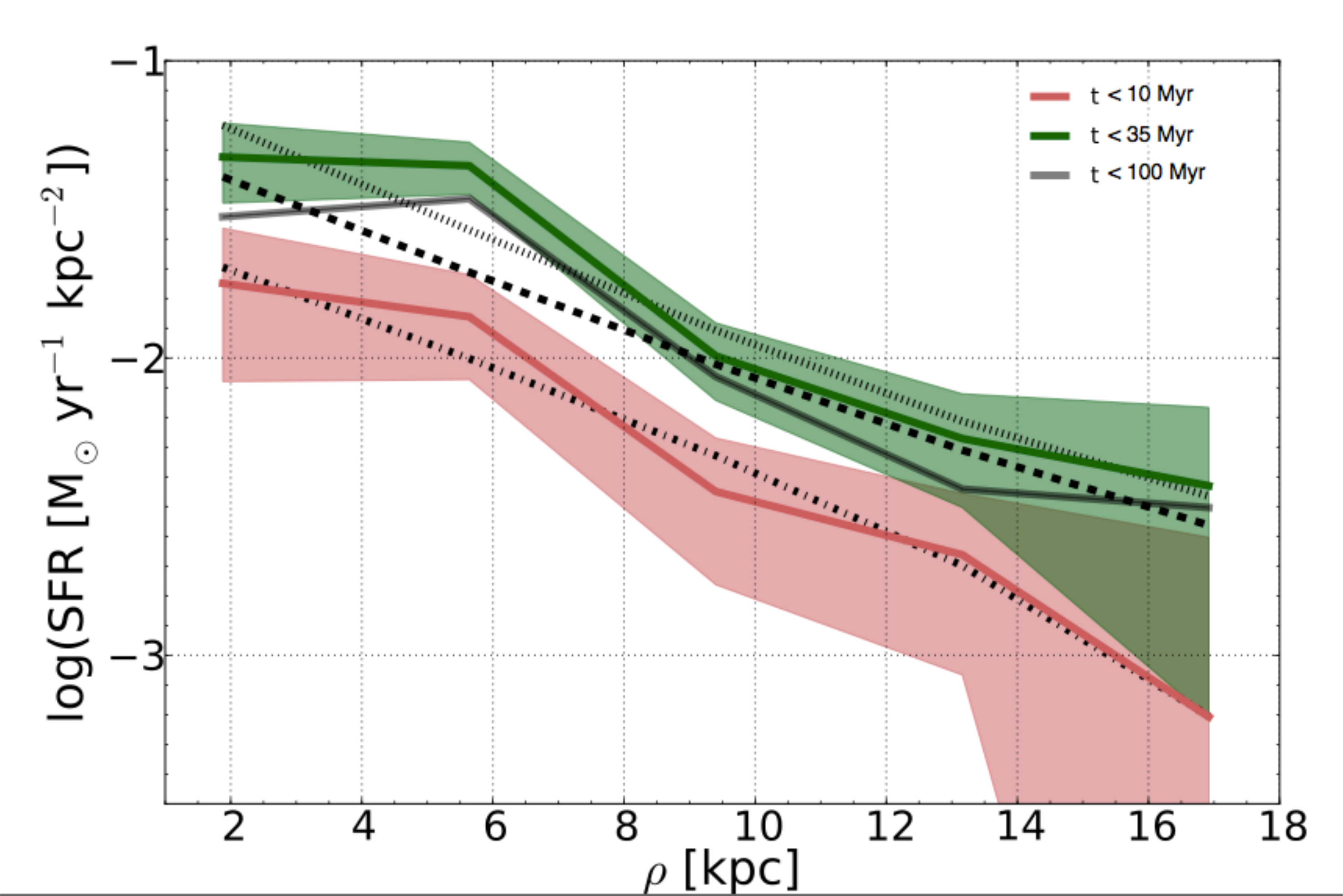} 

\caption{Mean SFRs as a function of radius for stars $<10$ Myr, $<35$ Myr, and $<100$ Myr.  The shaded regions surrounding the lines denote the $1\sigma$ error bars.  Exponential functions were fit to each of the radial SFR  profiles, with best-fit scale lengths (in kpc) of $4.65\pm0.21$ ($<10$ Myr; dot-dashed line), $6.38\pm0.26$ ($<35$ Myr; dashed line), and $6.30\pm0.15$ ($<100$ Myr; dotted line).}
\label{fig:radial_sfr}
\end{figure} 

\subsection{Radial Variations in the Star Formation History}  
FUV emission from the photospheres of young O and B stars traces stellar populations with ages $\lesssim35$ Myr, while the NUV is sensitive to older stellar populations $\lesssim100$ Myr.  Nebular H${\alpha}$ emission, on the other hand, provides an instantaneous glimpse of the most recent ($\lesssim10$ Myr) bursts of massive star formation.  Assuming a Kennicutt-Schmidt law \citep{Kennicutt:1998a}, the mean SFR for times $<10$, $<35$, and $<100$ Myr should correlate with the H$\alpha$, FUV, and NUV emission profiles, respectively.  Thus, we determine the mean SFR for times $<10$, $<35$, and $<100$ Myr and plot them as a function of radius in Figure~\ref{fig:radial_sfr}.  We then fit exponential functions to each of the radial SFR profiles and display their best-fit scale lengths on the left side of Table~\ref{tab:scalelengths}.  

To make the comparison between the radial SFR profiles and their corresponding emission profiles, we computed azimuthally averaged radial profiles, in $25\arcsec$ bins, using archival \textit{GALEX} \citep{Morrissey:2007} and H${\alpha}$ \citep{Hoopes:2001} images of M101 (Figure~\ref{fig:radial_profiles}).  In our determination of the radial profiles, we found that radius at which the UV profiles reach background sky values is at $\approx15\arcmin$.  The H$\alpha$ profile, on the other hand, reaches background at a radius of $10\arcmin$, which agrees well with observed H$\alpha$ truncation radius of $10.6\arcmin$ determined by \cite{Martin:2001}.  We then fit exponential functions to each of the radial emission profiles; their best-fit scale lengths are provided on the right side of Table~\ref{tab:scalelengths}.

\begin{figure}[hb!] 
\centering

\includegraphics[width=\columnwidth]{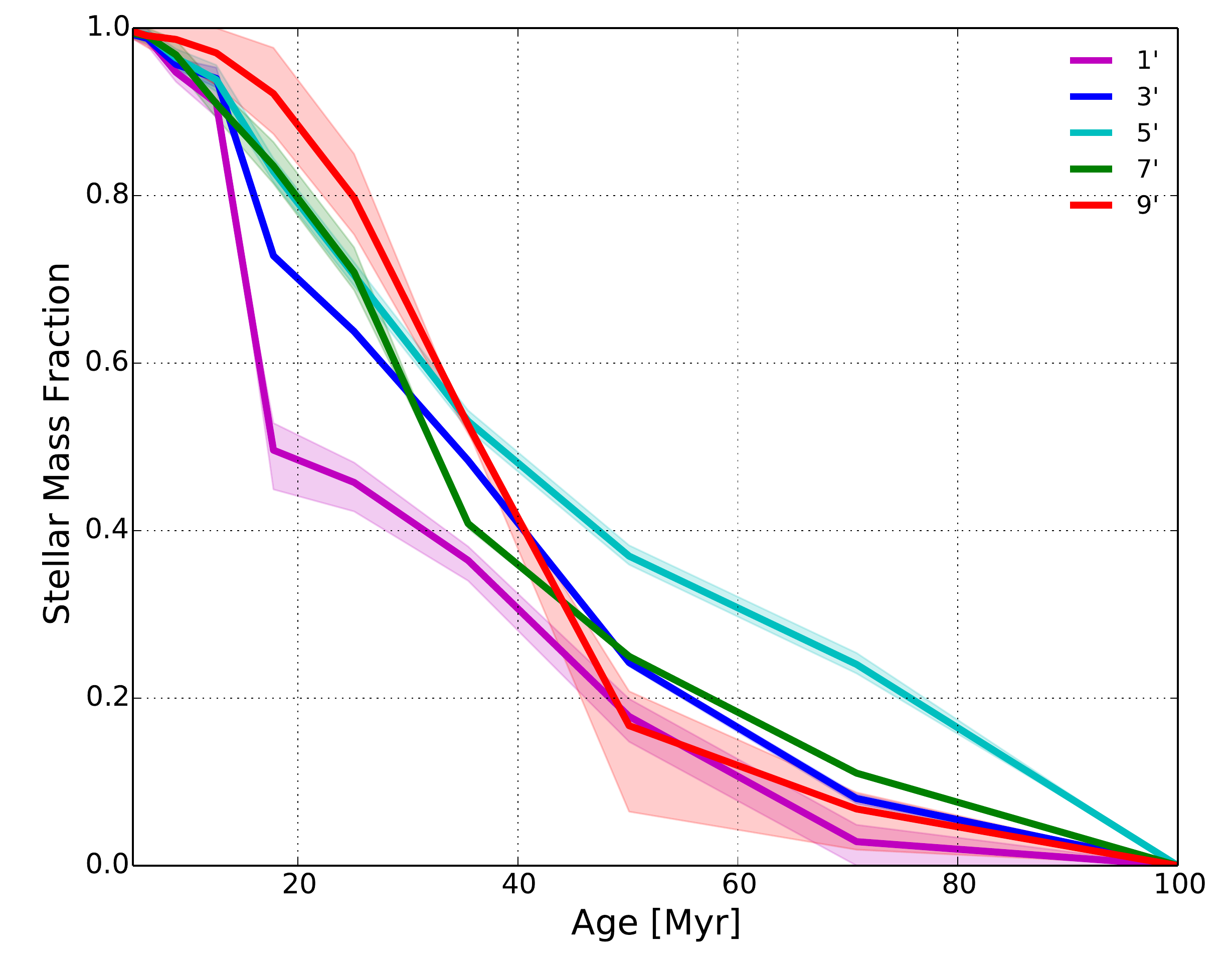} 

\caption{Cumulative SFH for each annulus.  The stellar mass fractions are with respect to the mass formed over the last 100 Myr.  The legend gives the radial center of each bin in arcminutes.  The shaded regions denote the 1$\sigma$ error bars.}
\label{fig:radialCSF}
\end{figure}

There is excellent agreement between the radial SFR profiles and their corresponding radial emission profiles suggesting that the spatial variations in the SFH are well correlated with the emission properties of M101.  Table~\ref{tab:scalelengths} indicates that the radial SFR profiles for stars $<35$ and $<100$ Myr old are considerably more extended than that of stars $<10$ Myr old.  In the inner regions, the ratio of the SFR in the last 35 Myr to the SFR in the last 10 Myr is approximately 2, whereas at 17 kpc it increases to approximately 10 (Figure~\ref{fig:radial_sfr}).  Since this ratio does not remain static, the relative fraction of stellar populations that are $<35$ Myr and $<10$ Myr likely changes dramatically with radius.

\begin{figure}[htb!]
\centering

\includegraphics[width=\columnwidth]{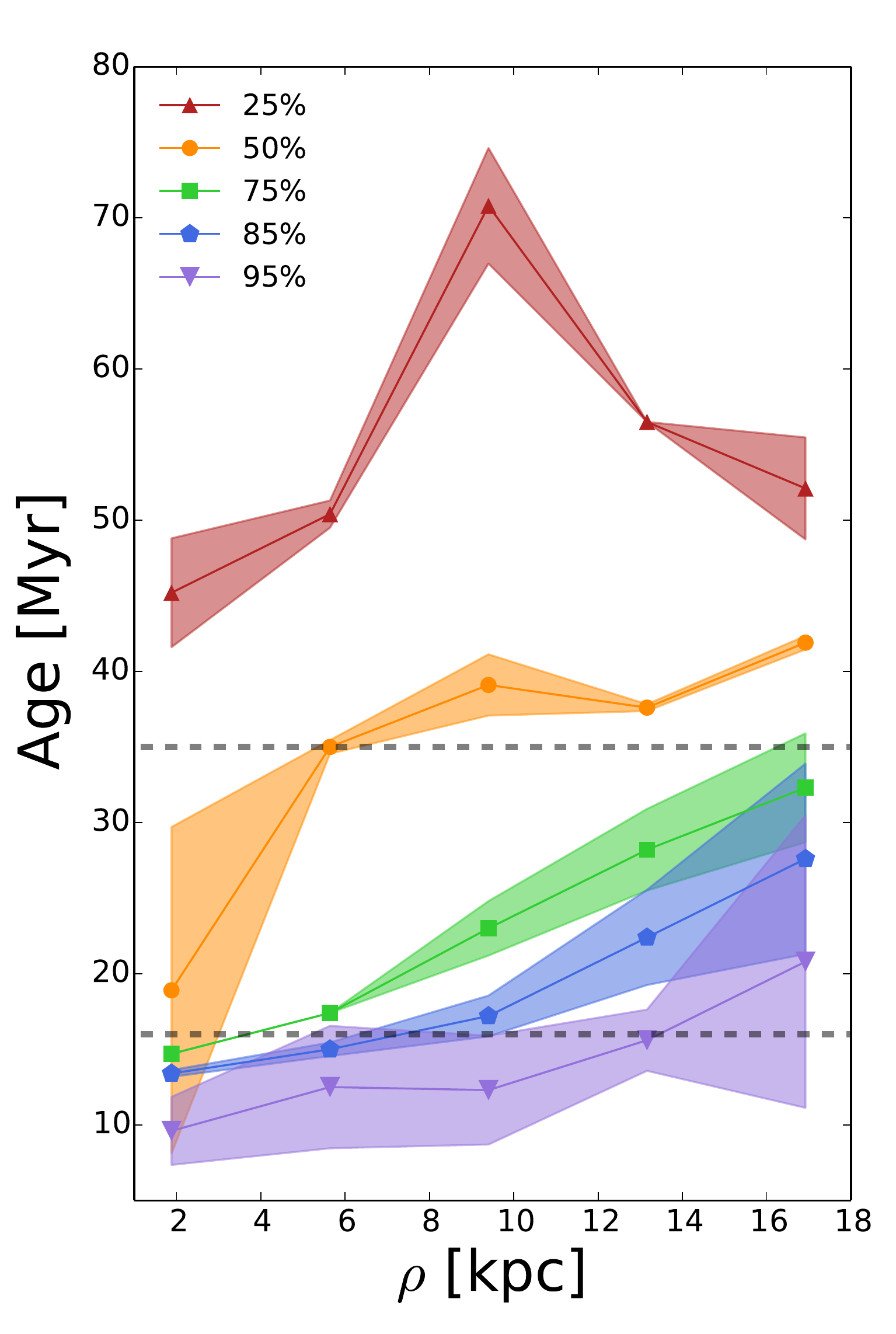} 

\caption{The times at which the $25\%, 50\%, 75\%, 85\%$ and $95\%$ stellar mass fractions are reached as a function of radius.  1$\sigma$ error bars are represented by the shaded regions.  The horizontal lines indicate the stellar population ages that contribute $99.9\%$ of the H$\alpha$ (16 Myr) and $80\%$ of the FUV (35 Myr) photons \citep{Gogarten:2009}.}
\label{fig:age_gradient}
\end{figure} 

To determine how the relative proportions of stars $<10$, $<35$, and $<100$ Myr old vary with radius, we compute the cumulative SFH (Figure~\ref{fig:radialCSF}).  The cumulative SFH shows how stellar mass has accumulated over time.  The dependent variable is the fraction of stars that are older than a specified age; one minus the fraction yields the proportion of stars that are younger.  By definition, $100\%$ of the stars are older than 4 Myr (our youngest isochrone), and similarly, $100\%$ of the stars are younger than 100 Myr.  Figure~\ref{fig:radialCSF} shows that the epoch of star formation occurring between $50-100$ Myr ago roughly accounts for $20\%-25\%$ of the 100 Myr stellar mass fraction.  For more recent times, we find that for a fixed age, the outer radii exhibit larger mass fractions than the inner radii, suggesting the mean stellar age increases with radius.  

To better demonstrate this trend, we determine the age at which the $25\%, 50\%, 75\%, 85\%$, and $95\%$ mass fraction are attained for each cumulative SFH (Figure~\ref{fig:age_gradient}).  With the exception of the annulus at $5\arcmin$ ($\rho \approx 9.5$ kpc), the age-radius relation for the $25\%$ mass fraction is relatively flat, with each annulus achieving $25\%$ around 50 Myr ago.  The $50\%, 75\%, 85\%$, and $95\%$ age-radius relations increase in age by approximately 10 Myr over the 17 kpc radial extent.  Spectral synthesis models of H$\alpha$, FUV, and NUV emission as a function of age indicate that $99.9\%$ of the H$\alpha$ photons, and $55\%$ of the UV photons, are emitted by stellar populations $<16$ Myr old \citep{Gogarten:2009}.  From Figure~\ref{fig:age_gradient}, we find that the fraction of stars that are $<16$ Myr old is $15\%-35\%$ in the inner radii compared to $<5\%$ at 17 kpc.  For comparison, stars $<35$ Myr old account for $\gtrsim50\%$ in the inner regions and $\sim25\%$ at 17 kpc.  Thus, the data are consistent with a stellar age gradient tending towards older stars in the outer disk.  If we assume that the observed trends in the SFH persist beyond 17 kpc, the fraction of stars $<10$ Myr old is unlikely to be sufficient to produce significant H$\alpha$ emission.  Therefore, the dearth of H$\alpha$ emission at large radii may be attributed to an increase in the mean stellar age with radius.  

While our data are consistent with an age gradient, we must consider the possiblity that stochastic effects due to the statistical sampling of the IMF, or a high mass truncation in the IMF, are resulting in the observed trends.  The data clearly indicate a radial gradient in the maximum stellar mass that has been interpreted as a gradient in stellar age: an interpretation that hinges upon the assumption of an invariant IMF.  If the IMF were statistically sampled such that high mass stars were less likely to form in regions of low SFR, or in low density regions there existed a rigid upper mass limit above which stars could not form, the lack of high mass stars could be incorrectly interpreted as older stellar populations.  As a result, our data are consistent with a radial age gradient that is sufficient to explain the discrepancies between the H$\alpha$ and UV radial emission profiles, however, we cannot conclusively eliminate alternative explanations.

\begin{figure}[htb!]
\centering

\includegraphics[width=\columnwidth]{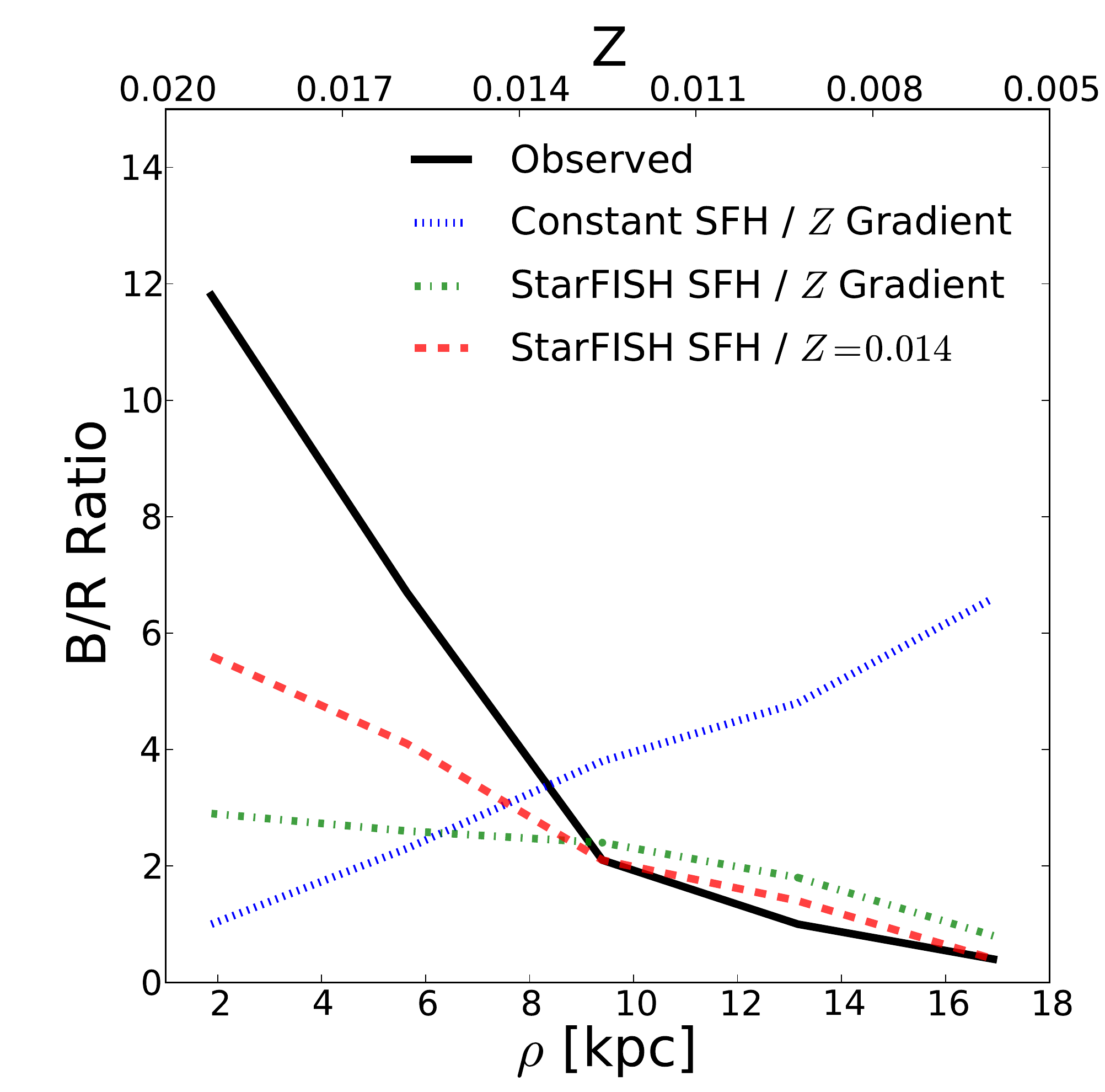} 

\caption{Modeled blue-to-red supergiant ratios. Model CMDs were drawn from a constant SFH with the observed metallicity gradient (dotted), the derived SFH and a metallicity gradient (dot dashed), and the derived SFH and no metallicity gradient (dashed).  The observed blue-to-red supergiant ratio is given as a solid line.}
\label{fig:br_ratio}
\end{figure} 

\subsection{Modeling the Blue-to-Red Supergiant Ratio}
The main candidates for the determination of the B/R ratio are population age, metallicity, and stellar rotation (see \cite{Meynet:2011} and references therein).  Observations dictate that the B/R ratio is an increasing function with metallicity (see \cite{Eggenberger:2002} and references therein).  It has also been shown that the B/R ratio sensitive to stellar population age \citep{Dohm-Palmer:2002, Vazquez:2007, McQuinn:2011} and the shape of the B/R ratio as a function of age is adequately modeled using non-rotating stellar isochrones for low metallicities ($Z < 0.009$).  In Paper I, we show that the B/R ratio decreases with radius and decreasing metallicity.  As we have demonstrated in the above sections, the stellar populations in M101 increase in age with radius.  Given the dependence of the B/R ratio on SFH, and the fact that there is no publically available set of rotating isochrones which span the range of metallicities in M101, the goal of this section is to use non-rotating stellar isochrones to model the B/R ratio in each annulus, including the effects of the SFH and metallicity gradient.  To do so, we: i) identify blue and red supergiants in our observed CMDs, ii) examine how our selection criteria affect the predicted B/R ratios and their variation with age and metallicity, and iii) model the B/R ratio in each annulus.  

In Paper I, we identified blue and red supergiants using color-color and color-magnitude criteria with a limiting absolute magnitude of M$_V < -5$.  Binning the stars into three luminosity bins, we found that the B/R ratio was a decreasing function with radius, dropping by nearly a factor of 15 in all three lumninosity bins.  For this study, we determine the B/R ratio in each annulus, for all stars brighter than M$_V = -5$.  In Figure~\ref{fig:br_ratio}, we show the observed B/R ratio as a function of radius and find a similar trend to that in Paper I.   

\begin{figure}[htb!]
\centering
\includegraphics[width=\columnwidth]{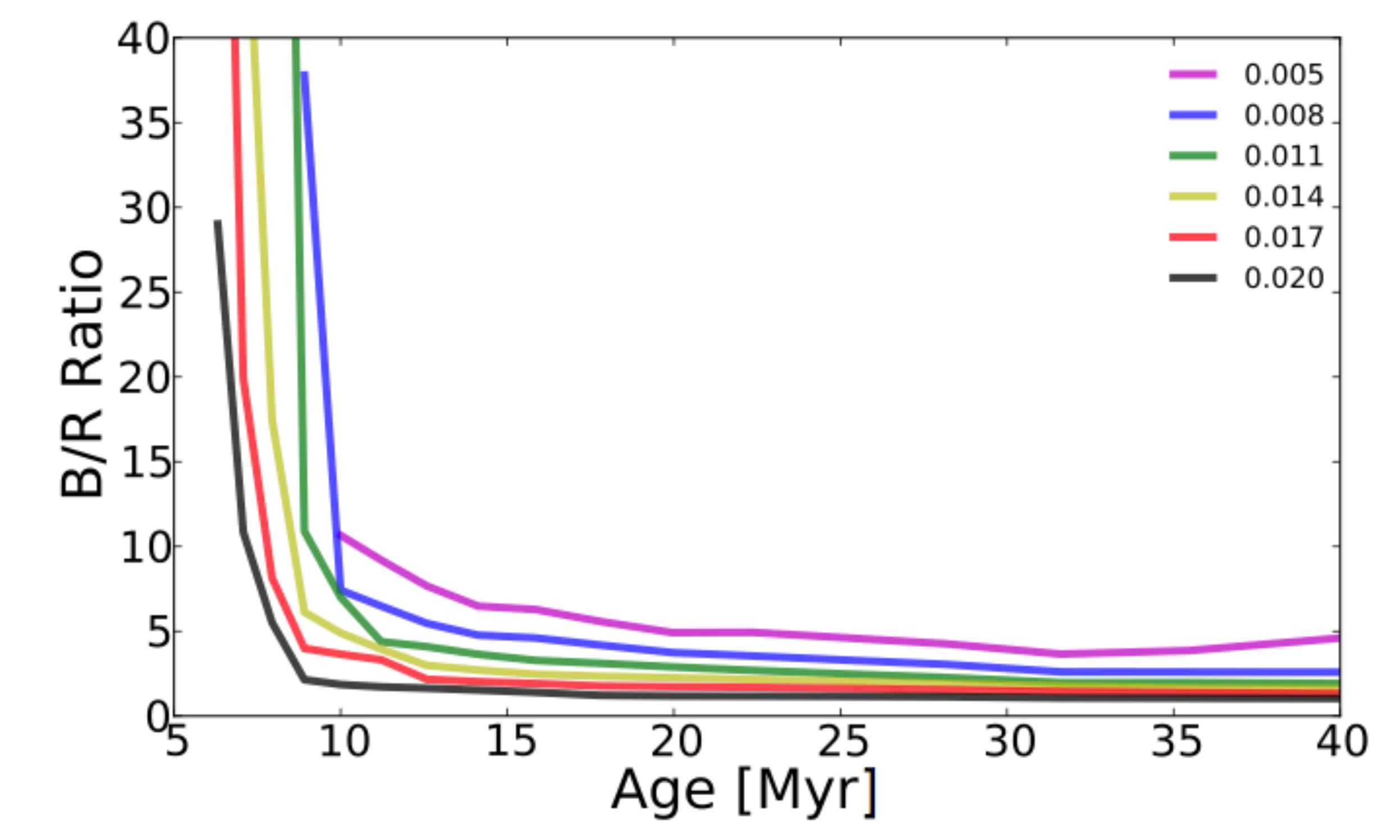} 
\caption{B/R ratio versus age for metallicities between $Z = 0.005$ and $Z = 0.020$.  Blue and red supergiants were identified using the selection criteria from Paper I. The youngest age at which the B/R ratio is calculable is when red supergiants first appear on the CMD.  Red supergiants appear as early as 7 Myr for $Z = 0.020$, and as late as 10 Myr for $Z = 0.005$.  One may equivalently interpret this as the upper mass limit at which red supergiants may exist increasing with metallicity.}
\label{fig:br_age}
\end{figure}

Next, we examine how the B/R ratios are expected to vary with age and metallicity when identified using our selection criteria.  We created synthetic single-age CMDs from the Padova stellar isochrones \citep{Marigo:2008} for the metallicities between $0.005-0.020$.  Each CMD includes $\sim200000$ stars drawn from an IMF with slope -2.35 \citep{Salpeter:1955}.  From each synthetic CMD, we photometrically identify blue and red supergiants using the selection criteria given in Paper I.  Our absolute magnitude cutoff of M$_V < -5$ corresponds to an age cutoff of $\sim40$ Myr \citep{Marigo:2008}, therefore we calculate the B/R ratio for ages $<40$ Myr old.  In our calculation, the youngest age at which the B/R ratio is calculable is set by the upper luminosity limit for red supergiants, which corresponds to an upper mass limit of $\sim40-50$ M$_{\odot}$ \citep{Humphreys:1979a, Humphreys:1994}.  We show the relationship between B/R ratio and age for all six metallicities in Figure~\ref{fig:br_age}.  We find that our selection criteria result in a B/R ratio that is initially very large ($>30$) followed by a precipitous decline, analogous to the age dependence of the B/R ratio shown in \cite{Vazquez:2007}.  Furthermore, the age at which the red supergiants first appear on the CMD is metallicity dependent, with red supergiants appearing as early as 7 Myr for $Z = 0.020$, and as late as 10 Myr for $Z = 0.005$ (one may equivalently interpret this as the upper mass limit at which red supergiants may exist increasing with metallicity).  While red supergiants appear earlier in high metallicity isochrones, the low metallicity isochrones predict, on average, higher B/R ratios for ages $>10$ Myr.  
 
To test the likelihood that variations in the SFH with radius are responsible for the change in the B/R ratio with radius, we model the B/R ratio under three conditions: i) a constant SFH and observed metallicity gradient, ii) the derived SFH and constant metallicity, and iii) the derived SFH and observed metallicity gradient.  To model the B/R ratio, we created model CMDs, of a single metallicity, for each annulus as the linear combination of single-age CMDs.  The number of stars in the model CMD contributed by each single age CMD was determined by the SFR multiplied by width of the time bin and divided by the average mass.  From our modeled CMDs, we identified blue and red supergiants more luminous than M$_V = -5$ using the  selection criteria from Paper I.  In Figure~\ref{fig:br_ratio}, we display the predicted B/R ratios and their dependence on radius for all three conditions.  Additionally, we display the metallicity gradient in M101 \citep{Kennicutt:2003} as the second x-axis on Figure~\ref{fig:br_ratio}.  

In our first condition, we created synthetic CMDs assuming a constant SFH over the last 100 Myr as well as the metallicity gradient from \cite{Kennicutt:2003} (dotted line).  After selecting blue and red supergiants from the synthetic CMDs, we plot the B/R ratio as a function of radius.  We find that using a constant SFH, the B/R ratio is an increasing function with radius and increases with decreasing metallicity.  This scenario illustrates the inability of current stellar evolution models to match observations.  

In our second condition, we remove the metallicity gradient by calculating the B/R ratios in each annulus assuming $Z = 0.014$.  In our calculation, we take the derived SFH into account.  This condition allows us to illustrate the effect that the observed age gradient has on the predicted B/R ratios in a single metallicity environment.  Figure~\ref{fig:br_ratio} indicates that the radial B/R ratio trend is reversed in the presence of the apparent age gradient and the absence of the metallicity gradient (dashed line) which is unsurprising based on how the B/R ratio is expected to depend on age and metallicity (Figure~\ref{fig:br_age}).  We find that for radii greater than 9 kpc, the modeled and observed B/R ratios are in good agreement; whereas inside 9 kpc they differ by as much as a factor of 2.  While the agreement is interesting, M101 is not a single metallicity environment and so we must include the metallicity gradient to properly model the B/R ratio in each annulus.

Lastly, in our third condition, we created synthetic CMDs assuming the best-fit SFH as well as the metallicity gradient.  With the addition of metallicity gradient, we find that the modeled B/R ratio trend (dot-dashed line) is considerably shallower than observations and at large radii (low metallicity), the models predict B/R ratios that are larger than observations by $\lesssim2$.  Studies of the B/R ratios in low metallicity galaxies have found  similar agreement between model predictions and observations \citep{Dohm-Palmer:2002, Ubeda:2007, McQuinn:2011}.  The largest discrepancies occur at small radii (high metallicity) where we find that the observed ratios are greater than the models predict by a factor of approximately 4.  At high metallicity, the non-rotating models predict much lower B/R ratios than the observations dictate and only after adding in stellar rotation do the models and observations agree \citep{Vazquez:2007}. Thus, the observed discrepancies between our modeled and observed B/R ratios may be primarily due to our use of non-rotating stellar isochrones.  However, at this time, we cannot incorporate rotation into our modeled B/R ratios since there are no published rotating isochrones which span the metallicity range in M101.

\section{Summary and Future Work} 
In this paper, we use 16 archival \textit{HST}/ACS fields to determine the radial SFH of M101 and its influence on the emission properties and stellar populations.  We derive the SFH from the resolved stellar populations in five $2\arcmin$ wide annuli.  Our main conclusions are the following: 

\begin{enumerate}
\item Binning the SFH into time frames corresponding to stellar populations traced by H$\alpha$ ($<10$ Myr), FUV ($<35$ Myr), and NUV ($<100$ Myr) emission, we determined radial SFR profiles.  We fit exponential functions to the radial SFR profiles and compared them to the radial emission profiles themselves.  We find that radial SFR profiles and the radial emission profiles have best-fit scale lengths that are in excellent agreement.  Our results showed that the $<35$ Myr radial SFR profile is considerably more extended than that of the $<10$ Myr radial SFR profile.  Examining the age at which the $50\%, 75\%, 85\%$, and $95\%$ mass fractions are attained, we find that the mass fraction for stars $<16$ Myr old is $15\%-35\%$ in the inner regions, compared to less than $5\%$ in the outer regions.  The mass fraction for stars $<35$ Myr old is greater than $50\%$ in the inner regions and greater than $25\%$ in the outer regions.  These findings are consistent with a radially increasing stellar age gradient which provides a natural explanation for the lack of H$\alpha$ emission at large radius.  However, our data cannot rule out alternative explanations such as statitical sampling of the IMF or a strongly truncated IMF.

\item We model the B/R ratio in our five annuli, examine the effects that a metallicity gradient and variable SFH have on the predicted ratios, and compare to the observed values. Holding the SFH constant, the predicted B/R ratios decrease with metallicity, a trend which is opposite to observations.  Including the radially variable SFH, the radial behavior of the B/R ratio mimics that of the observed values.  Moreover, we find that our modeled B/R ratios closely match the observed values at large radii (low metallicity) but are discrepant at small radii (high metallicity), which we attribute to our use of non-rotating stellar isochrones.
\end{enumerate}

This is the second paper in a series on the massive star population in M101.  Paper III will be a spectroscopic survey of the most luminous stars and their variability from the multi-epoch, multi-color imaging survey using the Large Binocular Telescope \citep{kochanek:2008}.

\acknowledgments Our research on massive stars is supported by the National Science Foundation AST-1109394 (R. Humphreys, P.I.).  All of the data presented in this paper were obtained from the Mikulski Archive for Space Telescopes (MAST). STScI is operated by the Association of Universities for Research in Astronomy, Inc., under NASA contract NAS5-26555. Support for MAST for non-HST data is provided by the NASA Office of Space Science via grant NNX09AF08G and by other grants and contracts.  We thank the anonymous referee for a careful reading and detailed report which helped to improve this paper signiﬁcantly.

Facilities: \facility{HST/ACS}

\newpage
\bibliography{ms.bbl}
\newpage
\begin{appendices}
\begin{figure*}[ht!]

\centering 
\subfigure{\includegraphics[width=0.18\textwidth]{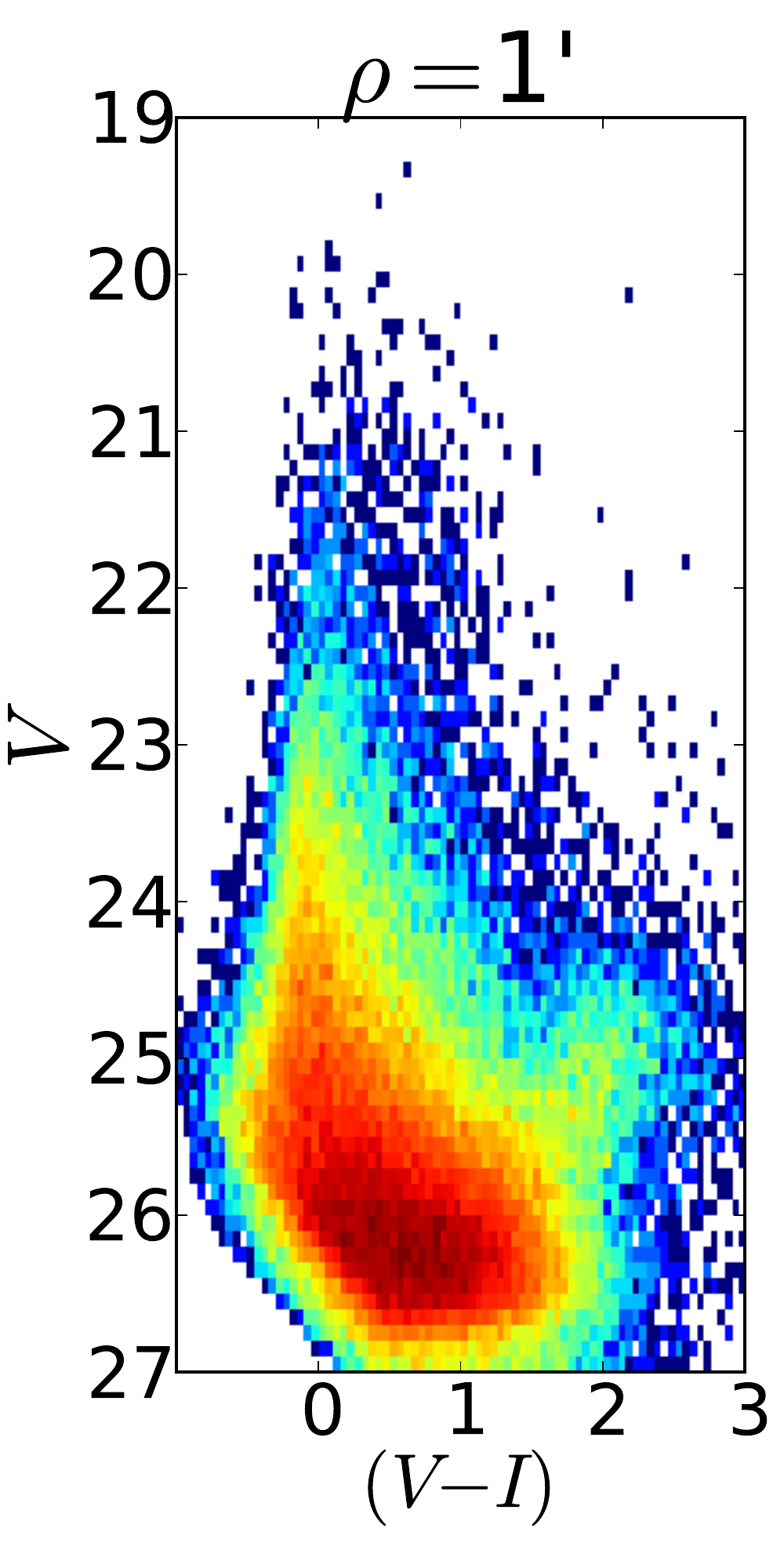}} 
\subfigure{\includegraphics[width=0.18\textwidth]{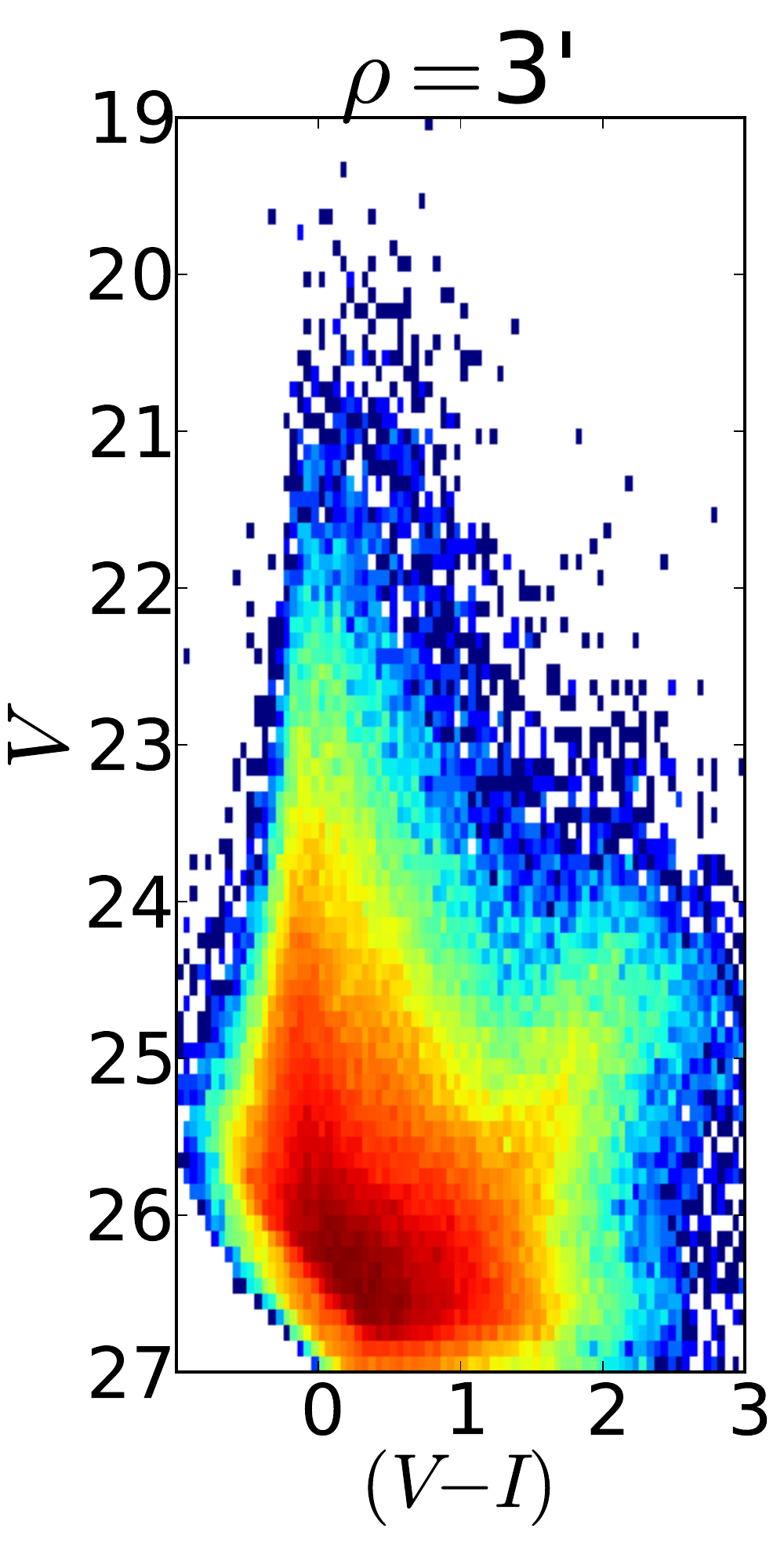}} 
\subfigure{\includegraphics[width=0.18\textwidth]{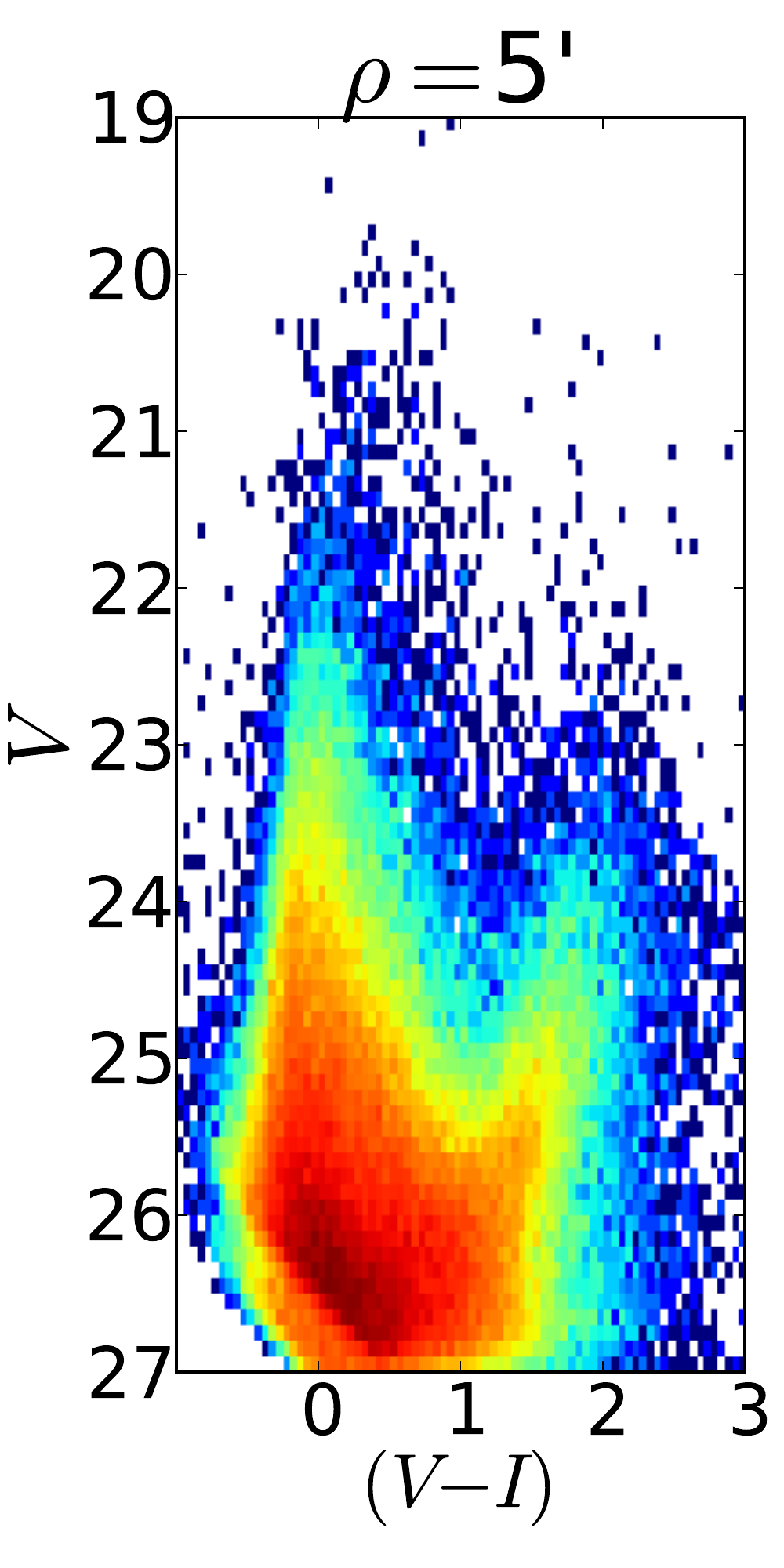}} 
\subfigure{\includegraphics[width=0.18\textwidth]{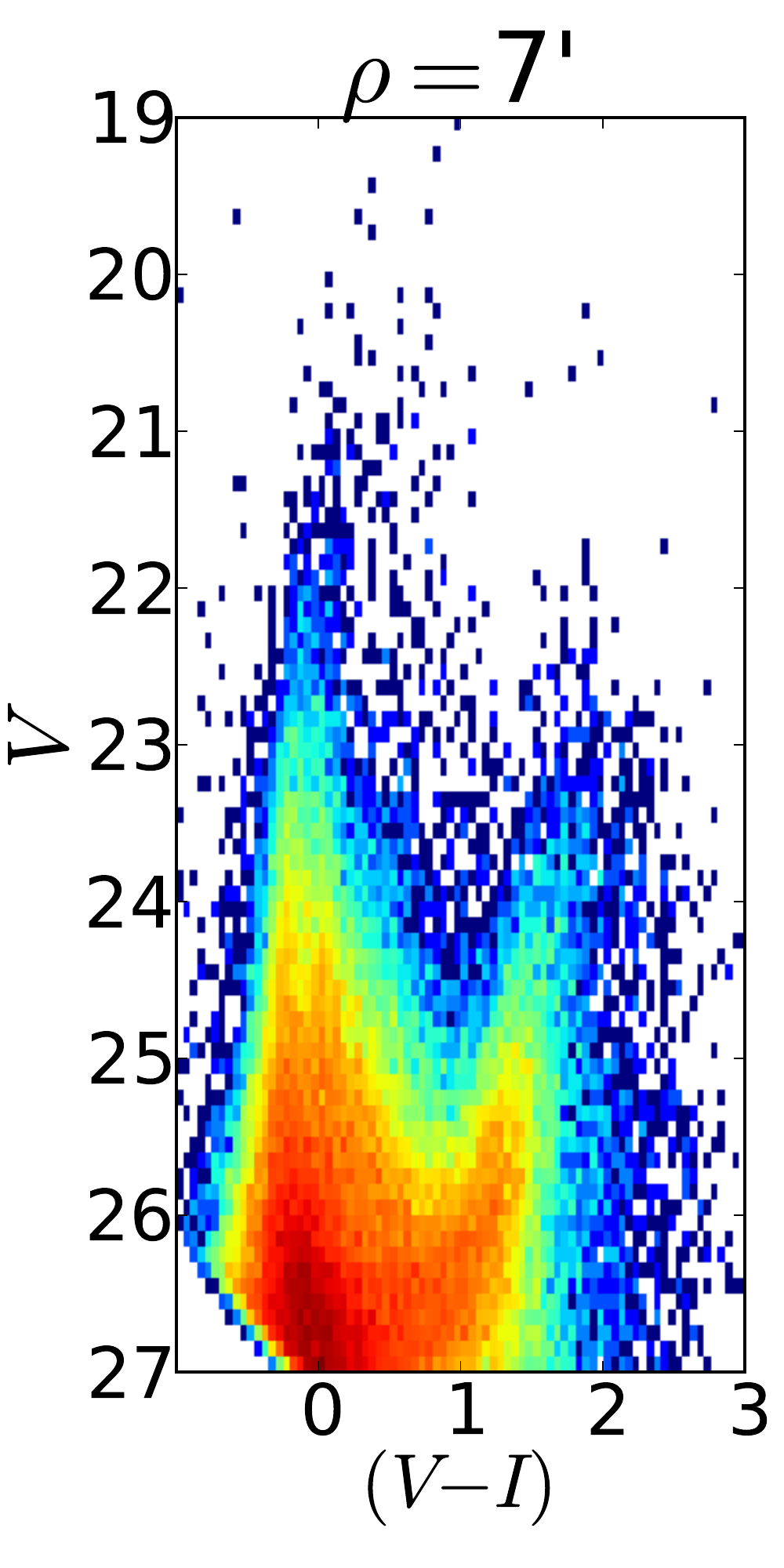}} 
\subfigure{\includegraphics[width=0.18\textwidth]{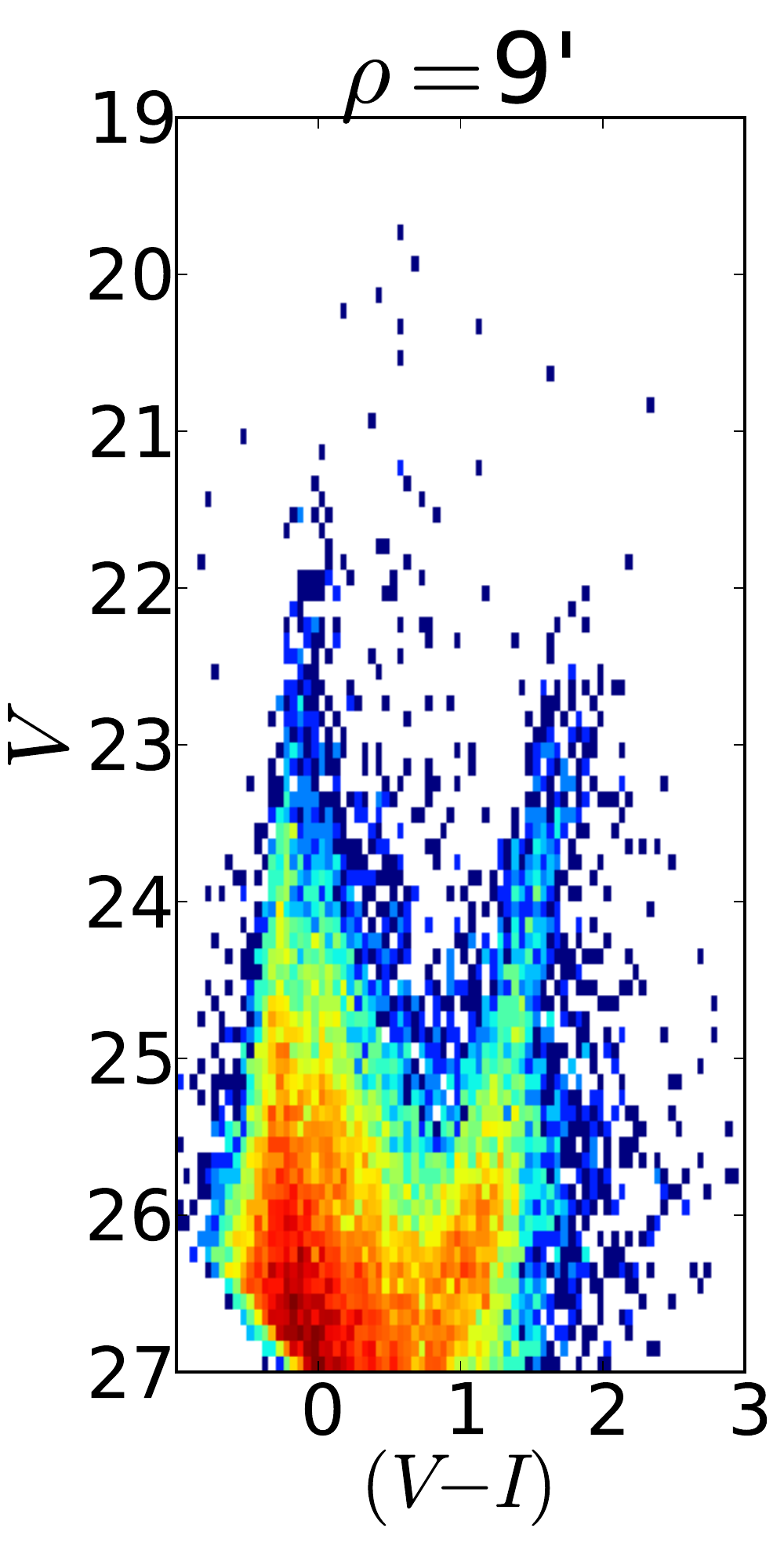}} 

\caption{$V$ vs. $(V-I)$ CMDs for each annulus.  The radial center is denoted at the top of each CMD.  The color map is linear in stellar density with lighter colors indicating higher densities.  A color version is available online with blue indicating low stellar density and dark red indicating high stellar density.} 
\label{fig:radial_Hess_VI}
\end{figure*} 
\end{appendices} 

\end{document}